\documentclass[11pt,a4paper]{article}

\usepackage{graphicx}% Include figure files

\begin{document}

\title{Unveiling the directional network behind financial statements data using volatility constraint correlation}
%Directionlity of Financial Accounting Variables by Volatility Constrained Correlation}

\author{Tomoshiro Ochiai \\Faculty of Social Information Studies, Otsuma Women's University, \\ 12 Sanban-cho,Chiyoda-ku,Tokyo 102-8357, Japan,
\\e-mail: ochiai@otsuma.ac.jp. \\ \\ Jose C. Nacher\\Department of Information Science, Faculty of Science, Toho University, \\Miyama 2-2-1, Funabashi, Chiba 274-8510, Japan, \\e-mail: nacher@is.sci.toho-u.ac.jp.}

\maketitle

\abstract{
Financial data, such as financial statements, contain valuable and critical information that may assist stakeholders and investors in optimizing their capital to maximize overall economic growth. Since there are many variables in financial statements, it is crucial to determine the causal relationships, that is, the directional influence between them in a structural way, as well as to understand the associated accounting mechanisms. However, the analysis of variable-to-variable relationships in financial information using standard correlation functions is not sufficient to unveil directionality. Here, we use the volatility constrained correlation (VC correlation) method to predict the directional relationship between two arbitrary variables. We apply the VC correlation method to five significant financial information variables (revenue, net income, operating income, own capital, and market capitalization) of 2321 firms listed on the Tokyo Stock Exchange over 28 years from 1990 to 2018. This study identifies which accounting variables are influential and which are susceptible. Our findings show that operating income is the most influential variable while market capitalization and revenue are the most susceptible variables. Surprisingly, the results differ from the existing intuitive understanding suggested by widely used investment strategy indicators, the price--earnings ratio and the price-to-book ratio, which report that net income and own capital are the most influential variables affecting market capitalization. This analysis may assist managers, stakeholders, and investors to improve financial management performance and optimize firms' financial strategies in future operations. 
}

\section{Introduction} 
The advent of information technology has made it possible to store and
classify a large amount of financial data in real time on an unprecedented scale \cite{Mantenga}. Financial information can be represented by a vast
variety of data types. However, one of the most abundant types of data
is multivariate time-series data, in which correlations among different
assets are often used to perform risk management analysis, determine investment strategies, and improve financial management.

The correlation structure in financial markets has been investigated
in many studies at different scales. Preis et al. analyzed 72 years of daily
closing prices of the 30 stocks of the Dow Jones Industrial Average and found that the average correlation among the stocks is linearly scaled with market stress at various time scales \cite{Preis}. Laloux et al. applied random matrix theory to understand the statistical structure of the correlation matrix of price changes and showed that there is a significant agreement between theoretical predictions and real data related to the density of eigenvalues \cite{Laloux}. This study raises serious questions about Markowitz's naive use of empirical correlation matrices for portfolio risk management. Plerou et al. also applied random matrix theory to analyze the cross-correlation matrix of price changes of the largest 1000 US stocks \cite{Plerou}. Their
findings showed that randomness cannot explain the universal and non-universal properties that enable
the identification of cross-correlations between stocks. Jiang and Zheng uncovered the positive and negative
subsector structures of Chinese, US, and global stock markets by
applying a random matrix theory analysis and taking into account
the sign of the components in the eigenvectors of the cross-correlation
matrix \cite{Jiang}. Cross-correlations between volume change and price change
of financial indexes were studied by \cite{Podobnik}. Long-range magnitude cross-correlations in real-world data from various fields, such as finance, have
also been studied intensively, using time-lag random matrix theory \cite{Podobnik2}. The
integration of random matrix theory and network methods was also
used to unveil correlation and network properties of 20 financial indexes \cite{Kumar}. The importance and novelty of physical methods applied to the study of financial cross-correlation analysis have been highlighted by Kwapien and Drozdz in their extensive review \cite{ Kwapien}. In particular, the application of random matrix theory to financial multivariate time series data deserved special attention. In this regard, several works have extended and further applied this statistical mechanics approach to the field of finance, namely \cite{Plerou2,Utsugi,Kwapien2,Toth,Biely,Kwapien3}.

However, there are many examples of datasets in finance in which
standard correlation metrics cannot detect the directionality of the
correlation (i.e., causality). To solve this directionality problem, Ochiai and Nacher \cite{vc1,vc2,vc3} introduced the volatility constrained correlation
(VC correlation) method, which enables us to determine the directionality of influence, a critical feature that cannot be determined by applying only standard correlation analysis techniques. The directionality of influence between the Japanese Nikkei 225 stock index and other financial markets is determined \cite{vc1}. VC correlation was applied to daytime and overnight returns, confirming the amplification of the negative correlation between them and consistency
with the time causality \cite{vc2}. Moreover, VC correlation was also used to analyze biological data and to predict gene regulatory interactions with high accuracy \cite{vc3}.

Meanwhile, financial statements are essential for decision
making, such as investment and mergers and acquisitions, by
shareholders, investors, and managers. Financial statements help improve the optimal
allocation of capital and promote whole economies. The VC correlation approach has not been applied to financial statement analysis, despite its importance in fraud analysis as well as profitability. Handayani et al. identified the absence of correlation of financial statement components as red
flags for detecting financial statement fraud \cite{Handayani}. Qualitative textual content in annual reports has been studied to predict fraud \cite{Goel}.
Ozdagoglu et al. classified financially correct and false statements
corresponding to Turkish firms listed on Borsa Istanbul using a decision tree, logistic regression, and artificial neural networks \cite{fraud}. Spathis developed a model for detecting the factors associated with false
financial statements using univariate and multivariate statistical techniques, such as logistic regression \cite{Spathis}. Sorkun and Toraman detected fraud in
e-ledgers through financial statements, using data-mining methods,
such as support vector machines, decision stump, M5P tree, J48 tree,
random forest, and decision tables \cite{Sorkun}.

Financial statements are closely linked to profitability and earnings prediction capability. This important subject has
been analyzed in many works using different techniques. For example,
Carstina et al. studied the correlation analysis of asset management and profitability indicators \cite{Carstina}. Ou and Penman studied the
connection of a large set of financial statement variables with the direction of 1-year-ahead earnings changes \cite{Ou}. Ishibashi et al. verified
the applicability of variable selection using data-mining techniques for
financial statement data \cite{Ishibashi}. Lokanan et al. evaluated the possibility
of rating the creditworthiness of a firm's quarterly financial report
using a dynamic anomaly detection method in the case of Vietnamese
listed firms \cite{Lokanan}. Andres et al. applied machine-learning algorithms
to study business profitability by preference learning \cite{Profitability}.
Wimmer and Rada applied information technology, such as decision trees and genetic algorithms, to fundamental financial statement data to predict
the direction of market capitalization \cite{Wimmer}.  
Yan and Zheng constructed fundamental signals from financial statements and used a bootstrap approach to evaluate the impact of data mining on fundamental-based anomalies, finding that they are better explained by mispricing \cite{Yan}.

 Although standard data analyses may reveal positive correlation
tendencies among most of these accounting values, the causality between
them has not been sufficiently investigated and clarified from a data
science viewpoint. The causality or directionality of influence between
them is difficult to determine using standard methods, such as standard
correlation coefficients. In this work, we determine the direction of influence among the
five primary financial accounting data (revenue, operating income, net
income, own capital, and market capitalization) using the data-driven
VC correlation technique and accounting data of 2321 firms listed from 1990 to 2018 on the Tokyo Stock Exchange, a standardized equities market in Japan. The observed directionality network of accounting variables enables us to observe the direct influence between accounting variables for the first time, suggesting future investment strategies and financial operations management. Our findings reveal not only that operating income is the most influential variable but also that market capitalization and revenue are the most susceptible variables. Interestingly, these predictions differ from
the conventional understanding suggested by widely used investment strategy indicators, the price-earnings ratio and price-to-book ratio, which posit net income and own capital as the most influential variables on market capitalization. Therefore, our novel analysis may assist managers, stakeholders, and investors to improve financial strategies in future operations, and also may be useful in related
financial engineering areas, such as fraud predictions and profitability
forecasts.

\section{Data}

The Tokyo Stock Exchange in Japan is the world's third largest exchange by market capitalization after the New York Stock Exchange and The Nasdaq Stock Market. Although there are many research works on the US market, there is also a necessity to investigate the Asian market which is emerging in the globe. In this study, we used the financial data of the Tokyo Stock Exchange, the largest exchange market in Asia, to study the directionality between financial data.

We use the five primary annual accounting data (revenue, operating income, net income, own capital, market capitalization) of 2321 firms listed on the Tokyo Stock Exchange over 28 years from 1990 to 2018. We exclude accounting data of banks, securities, non-life insurance, and life insurance companies, whose accounting data structure differs from that of general business companies.

\section{Methods}

\subsection{Metrics}
Let $r^c(t)$, $i^c(t)$, $p^c(t)$, $o^c(t)$, and $m^c(t)$ be revenue, net income, operating income, own capital, and market capitalization, respectively, of company $c$ in year $t$.

We define the rate of change of the five accounting variables (revenue, income, operating income, own capital, and market capitalization). The rate of change of revenue, own capital, and market capitalization are defined as follows:

\begin{eqnarray}
R_r^c(t)&=&\frac{r^c(t+1)-r^c(t)}{r^c(t)},\\
R_o^c(t)&=&\frac{o^c(t+1)-o^c(t)}{o^c(t)},\\
R_m^c(t)&=&\frac{m^c(t+1)-m^c(t)}{m^c(t)},
\end{eqnarray}
respectively. Similarly, the rate of change of net income and operating income are defined by
\begin{eqnarray}
R_i^c(t)&=&\frac{i^c(t+1)-i^c(t)}{r^c(t)},\\
R_p^c(t)&=&\frac{p^c(t+1)-p^c(t)}{r^c(t)},
\end{eqnarray}
respectively. Here, for the definition of $R_i^c(t)$ and $R_p^c(t)$, we use revenue $r^c(t)$ as the denominator instead of $i^c(t)$ and $p^c(t)$ in order to normalize the rate of change properly. The reason is as follows. If we use net income $i^c(t)$ or operating income $p^c(t)$ as the denominator, they could have a negative value or an exceedingly small value. In such a case,
the rate of change of income and operating income could take an extreme value, which would be difficult to handle.

\subsection{Standard correlation}
Let $R_s^c(t)$ be the change rate of accounting variable $s$ in year $t$ for company $c$ defined in the previous subsection. Here, $s$ represents $r$, $i$, $p$, $o$, or $m$ (revenue, net income, operating income, own capital, or market capitalization, respectively).

The average and standard deviation of $R_s^c(t)$ for a given period $[t_i,t_f]$ are given by
\begin{eqnarray}
&&E(R_s^c)=\frac{1}{(t_f-t_i)}\sum_{t_i \le t < t_f}R_s^c(t),\\
&&\sigma(R_s^c)=\sqrt{\frac{1}{(t_f-t_i)}\sum_{t_i \le t < t_f}(R_s^c(t)-E(R_s^c))^2}.
\end{eqnarray}

For a paired data set $\{(R_s^c(t),R_{s^\prime}^c(t)\}~(t_i \le t < t_f)$, the standard correlation coefficient (Pearson correlation coefficient) is given by
\begin{eqnarray}
C^c{[s,s^\prime]}=C(R_s^c,R_{s^\prime}^c)=\frac{1}{(t_f-t_i)}\sum_{t_i \le t < t_f}\frac{(R_s^c(t)-E(R_s^c))}{\sigma(R_s^c)}\frac{(R_{s^\prime}^c(t)-E(R_{s^\prime}^c))}{\sigma(R_{s^\prime}^c)}.
\end{eqnarray}

Let $E_C$ be the average of the correlation coefficient for all firms, and $\sigma_C$ be the standard deviation of correlation as follows:
\begin{eqnarray}
&&E_C[s,s^\prime]=E(C^c{[s,s^\prime]})=\frac{1}{N}\sum_{c}C^c{[s,s^\prime]},\\
&&\sigma_C[s,s^\prime]=\sigma(C^c{[s,s^\prime]})=\sqrt{\frac{1}{N}\sum_{c}(C^c{[s,s^\prime]}-E(C^c[s,s^\prime])^2},\nonumber\\
\end{eqnarray}
where $N$ is the total number of firms.

\subsection{Volatility constrained correlation}
In this subsection, we introduce VC correlation following \cite{vc1}. This metric enables us to amplify the correlation and identify the directionality of the influence.

Let $\Omega$ be a subset of all time points $\{t|t_i \le t < t_f\}$. We define the expectation value, standard deviation, and correlation coefficient, where the data points are constrained to the subset $\Omega$ as follows:
\begin{eqnarray}
&&E(R_s^c|\Omega)=\frac{1}{\#\Omega}\sum_{t\in\Omega}R_s^c(t),\\
&&\sigma(R_s^c|\Omega)=\sqrt{\frac{1}{\#\Omega}\sum_{t\in\Omega}(R_s^i(t)-E(R_s^c|\Omega))^2},\\
&&C(R_s^c,R_{s^\prime}^c|\Omega)=\frac{1}{\#\Omega}\sum_{t\in\Omega}\frac{(R_s^c(t)-E(R_s^c|\Omega))}{\sigma(R_s^c|\Omega)}\frac{(R_{s^\prime}^c(t)-E(R_{s^\prime}^c|\Omega))}{\sigma(R_{s^\prime}^c|\Omega))},\nonumber\\
&&
\end{eqnarray}
where $\#\Omega$ denotes the number of elements of $\Omega$.

As a special case of $\Omega$, we set
\begin{eqnarray}
\Omega_{R_s^c}=\{t\in [t_i,t_f] ~|~ h\sigma(R_s^c)\le |R_s^c(t)-E(R_s^c)|\},
\end{eqnarray}
where $h$ is the cutoff threshold, and in the following, we set $h=0.2$ so that we have enough data points to calculate the VC correlation. Later in Section \ref{sec:threshold}, we will discuss how the results change by varying the threshold $h$. For a pair of two accounting variables $s$ and $s^\prime$, we define the VC correlation between them as follows:
\begin{eqnarray} 
F^c{[s,s^\prime]}=C(R_s^c,R_{s^\prime}^c|\Omega_{R_s^c}),\\
F^c{[s^\prime,s]}=C(R_{s^\prime}^c,R_ s^c|\Omega_{R_{s^\prime}^c}).
\label{eqn:F}
\end{eqnarray}

Here, we call $F^c{[s,s^\prime]}$ and $F^c{[s^\prime,s]}$ the VC correlation between the accounting variables $s$ and $s^\prime$ of company $c$. If $F^c{[s,s^\prime]}$ is greater than $F^c{[s^\prime,s]}$, then the accounting variable $s$ has more influence on the accounting variable $s^\prime$ than in the opposite direction (i.e., accounting variable $s^\prime$ is more susceptible to accounting variable $s$).

We define the difference in VC correlations for the accounting variables $s$ and $s^\prime$ in company $c$ as follows:
\begin{eqnarray}
\Delta F^c{[s,s^\prime]}=F^c{[s,s^\prime]}-F^c{[s^\prime,s]}.
\end{eqnarray}

On the one hand, if $\Delta F^c{[s,s^\prime]}$ is positive, then the directionality of influence goes from $s$ to $s^\prime$. On the other hand, if $\Delta F^c{[s,s^\prime]}$ is negative, then the directionality of influence is from $s^\prime$ to $s$.

The expectation value and standard deviation of the difference in VC correlations $\Delta F^c{[s,s^\prime]}$ are given by
\begin{eqnarray}
&&E_{\Delta F}{[s,s^\prime]}=E(\Delta F^c{[s,s^\prime]}) = \frac{1}{N^\prime}\sum_{c \in M[s,s^\prime]}\Delta F^c{[s,s^\prime]},\\
&&\sigma_{\Delta F}{[s,s^\prime]}=\sigma(\Delta F^c{[s,s^\prime]}) =\sqrt{\frac{1}{N^\prime}\sum_{c \in M[s,s^\prime]} (\Delta F^c{[s,s^\prime]}-E(\Delta F^c{[s,s^\prime]}) )^2},\nonumber
\\
\end{eqnarray}
respectively. 

Here, 
\begin{eqnarray}
M[s,s^\prime]=\{c ~|~ C^c{[s,s^\prime]}>0,  (t_f-t_i)/2 \le \# \Omega_{R_s^c},  (t_f-t_i)/2 \le \# \Omega_{R_{s^\prime}^c} \},
\end{eqnarray}
where $\# \Omega_{R_s^c}$ and $\# \Omega_{R_{s^\prime}^c}$ denote the number of elements of $\Omega_{R_s^c}$ and $\Omega_{R_{s^\prime}^c}$ respectively, and $N^\prime$ indicates the number of elements of  $M[s,s^\prime]$.

We examine the statistical significance of the directionality between the two accounting variables by computing the p-value when the null hypothesis is that the mean value of $\Delta F^c{[s,s^\prime]}$ is zero. The threshold for statistical significance is that the p-value is less than $5\times 10^{-4} (=p_0)$ . Here, we evaluate the p-value from the Z-score computed by $Z=E_{\Delta F}{[s,s^\prime]}/(\sigma_{\Delta F}{[s,s^\prime]}\sqrt{N^\prime})$. The results are shown in detail in the next section.

\section{Results}

\subsection{VC correlation analysis predictions}
\label{sec:VC correlation analysis predictions}

\begin{figure}[htbp]

%\begin{center}

\hspace*{-4cm}
\includegraphics[scale=0.6]{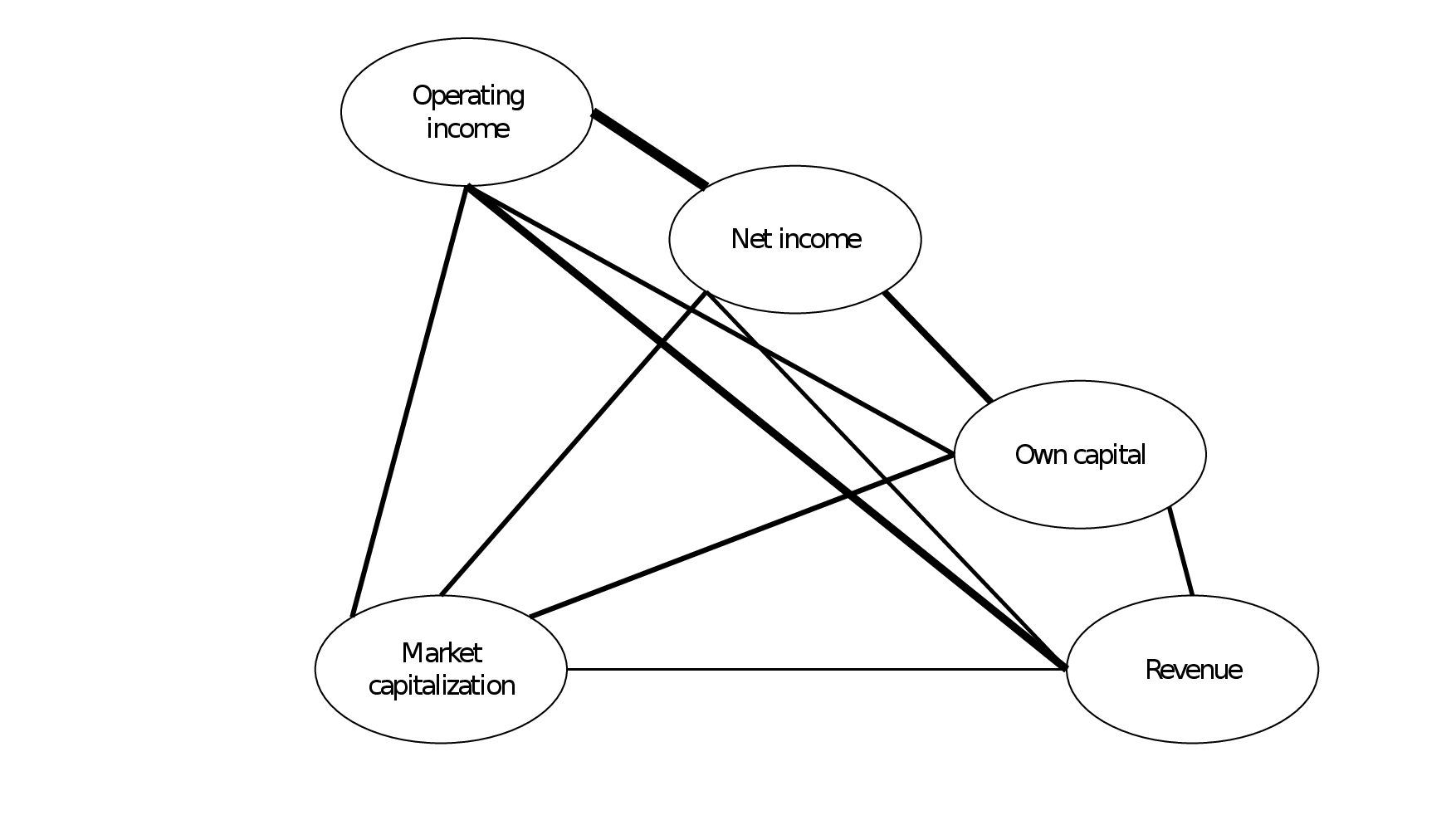}

%\put(-200,150){(b)}

%\end{center}

\caption{Correlation network. Thickness is linearly determined by the expectation value of correlation coefficient $E_C[s,s^\prime]$ between accounting variables $s$ and $s^\prime$.}

\label{fig:Correlation network}

\end{figure}

In Table \ref{tab: correlation coefficients}, we show the results for the average of correlation $E_C{[s,s^\prime]}$, the standard deviation of correlation $\sigma_C{[s,s^\prime]}$, and the number of firms $N^\prime$ for VC correlation computation for each paired accounting variable $[s,s^\prime]$. Table \ref{tab: correlation coefficients} shows that all the averages of the correlation coefficient $E_C[s,s^\prime]$ are positive, indicating that all five significant accounting variables are positively correlated. From these results, we plot the undirected network for the correlation relationship between the five accounting variables, as shown in Fig. \ref{fig:Correlation network}.

The standard correlation analysis shown in Fig. \ref{fig:Correlation network} provides no information about the directionality between the accounting variables. By contrast, the VC correlation analysis enables us to observe the directionality between accounting variable pairs. In Table \ref{tab: VC correlation}, we show the expectation values of the difference of VC correlation $E_{\Delta F}{[s,s^\prime]}$, the standard deviations of the difference of VC correlation $\sigma_{\Delta F}{[s,s^\prime]}$, the directionality from one variable to another shown by arrow symbols (i.e., ``$\rightarrow$'' or ``$\leftarrow$''), and p-values as the statistical significance of the predicted directionality.

From the expectation value $E_{\Delta F}{[s,s^\prime]}$ and the standard deviation $\sigma_{\Delta F}{[s,s^\prime]}$ of the difference of the VC correlation shown in Table \ref{tab: VC correlation},  we can determine the directionality (i.e., ``$\rightarrow$'' or ``$\leftarrow$'' in Table \ref{tab: VC correlation}) as follows. If $E_{\Delta F}{[s,s^\prime]}$ is positive (resp. negative) and the p-value is less than $p_0$, then the directionality is from the first item $s$ to the second item $s^\prime$ (resp. from $s^\prime$ to $s$) with statistical significance. If the p-value is more than $p_0$, there is no statistical significance, and we cannot say anything about directionality. In other words, when there is statistical significance, $E_{\Delta F}{[s,s^\prime]}$ is an indicator of the directionality between the accounting variables $s$ and $s^\prime$. We show the directionality in the right-most column in Table \ref{tab: VC correlation}.

As an example, we explain the first row of Table \ref{tab: VC correlation}. 
This row shows the result for the pair of net income(i) and market capitalization(m). The first variable $s$ is net income(i), and the second variable $s^\prime$ is market capitalization(m). Because $E_{\Delta F}[i,m]$ of this pair $[i,m]$ is positive ($E_{\Delta F}[i,m]=0.0110$), the directionality of the pair is from net income(i) to market capitalization(m) with very high statistical significance (p-value is $6.49\times10^{-9}$. This directionality is indicated by the right arrow $(\rightarrow)$ in the last column of the first row of Table \ref{tab: VC correlation}.

\begin{figure}[htbp]

\begin{tabular}{c}

\begin{minipage}{0.7\hsize}

\begin{center}

\includegraphics[scale=0.4]{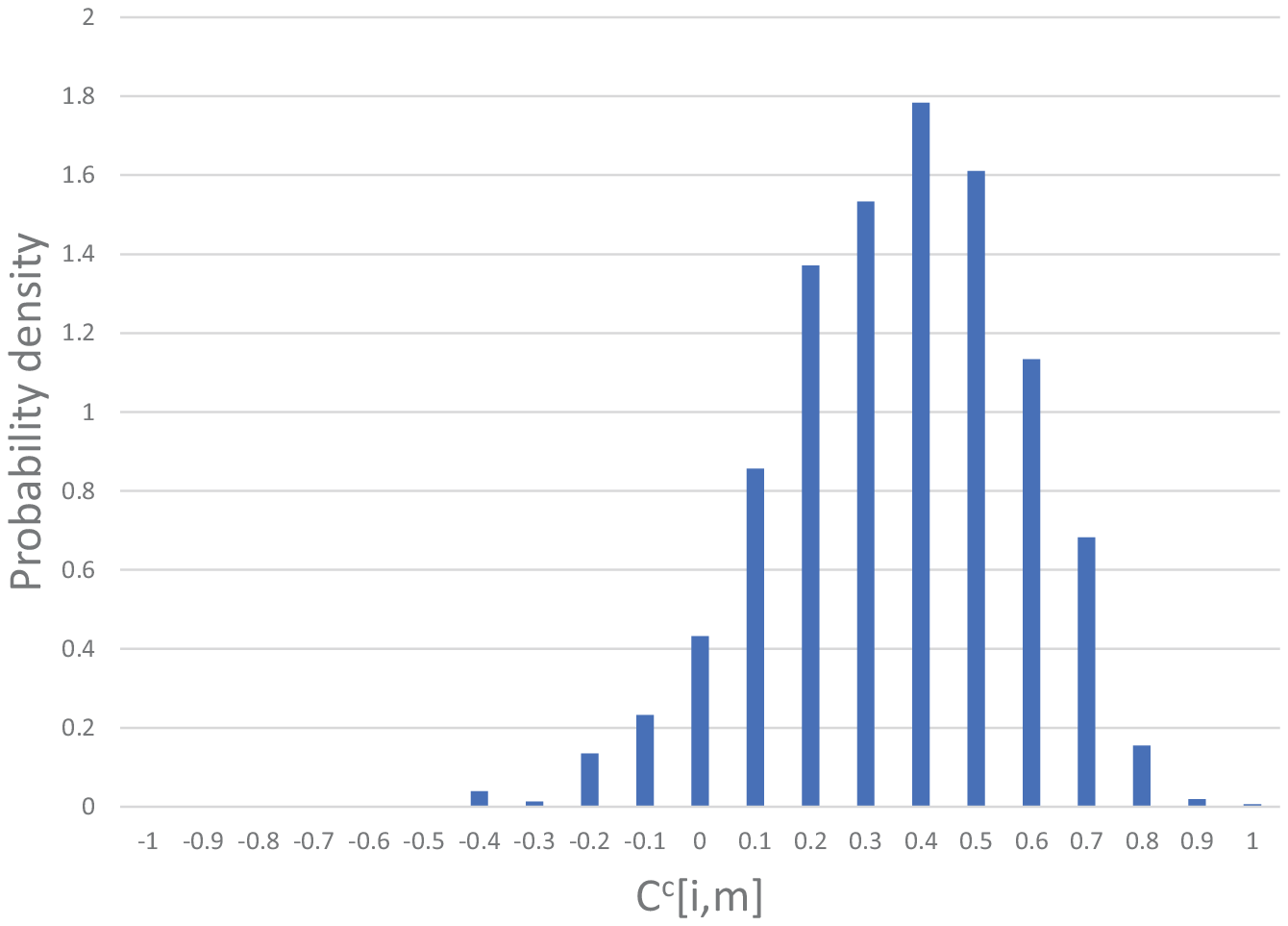}
\put(-300,220){(a)}

\end{center}

%\caption{}

%\label{}

\end{minipage}

\\

\begin{minipage}{0.7\hsize}

\begin{center}

\includegraphics[scale=0.4]{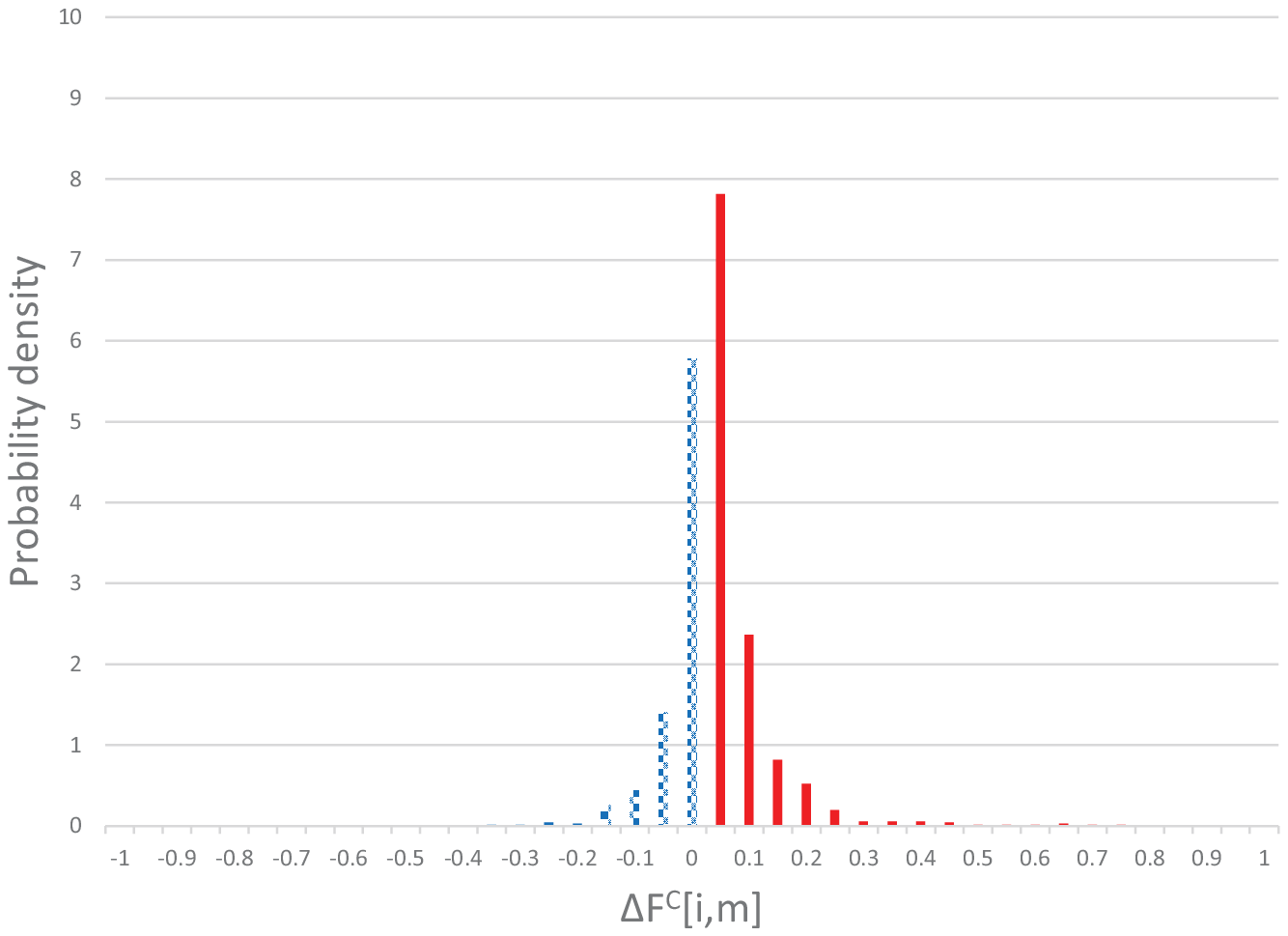}
\put(-300,220){(b)}

\end{center}

%\caption{}

%\label{}

\end{minipage}

\end{tabular}

\caption{(a) Probability density of the correlation coefficients between net income and market capitalization $C^c{[i,m]}$. (b) Probability density of the difference in VC correlations between net income and market capitalization $\Delta F^c{[i,m]}$. In (b), we observe asymmetry with respect to zero, which implies directionality from net income to market capitalization with enough statistical significance (p-value is $6.49\times10^{-9}$). See the first line in Table \ref{tab: VC correlation}. In both (a) and (b), the probability density is computed by normalizing the distribution.}
\label{fig:distribution assymetry}

\end{figure}

\begin{figure}[htbp]

\begin{tabular}{c}

\begin{minipage}{0.7\hsize}

\begin{center}

\includegraphics[scale=0.4]{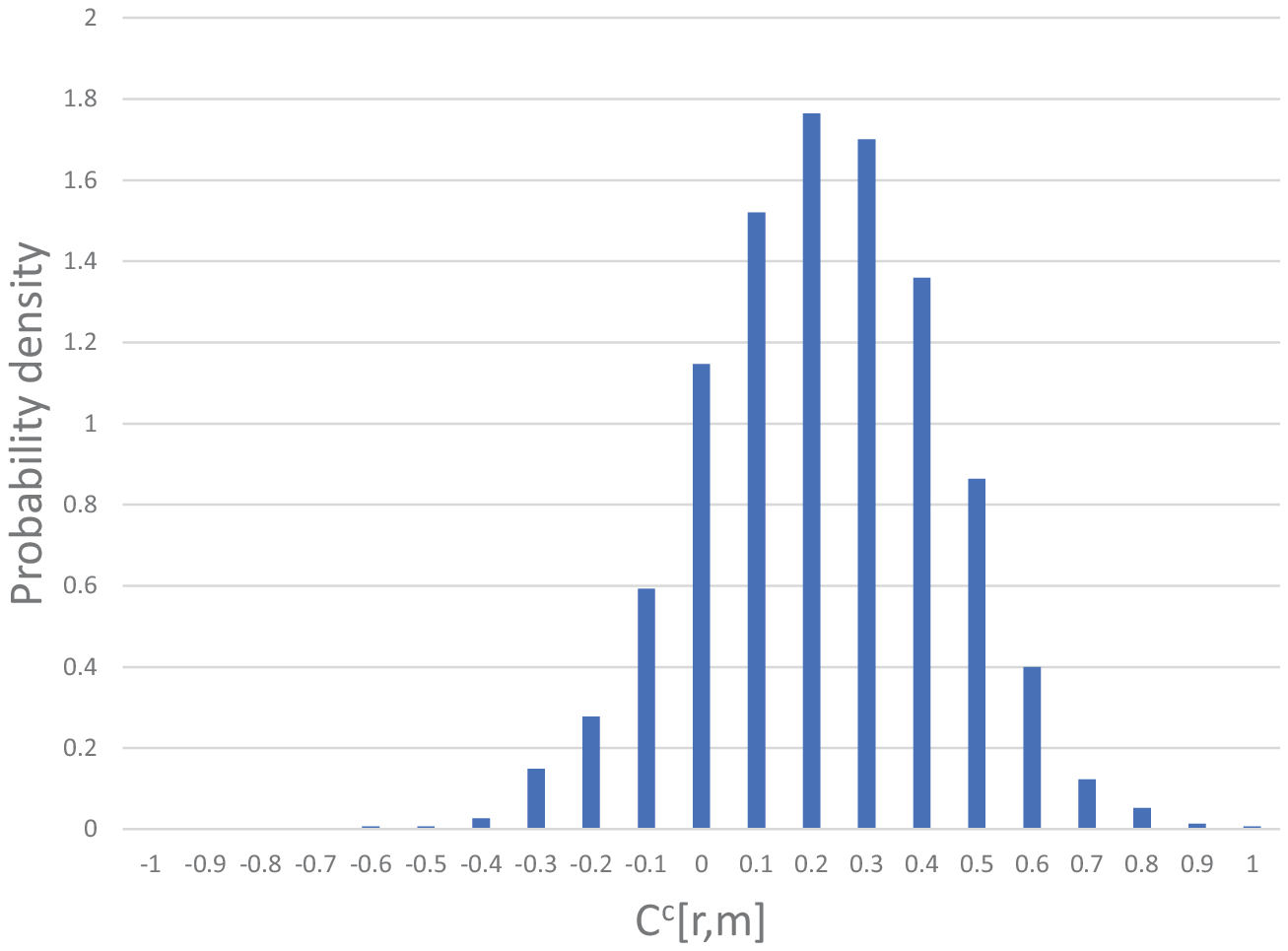}
\put(-300,220){(a)}

\end{center}

%\caption{}

%\label{}

\end{minipage}

\\

\begin{minipage}{0.7\hsize}

\begin{center}

\includegraphics[scale=0.4]{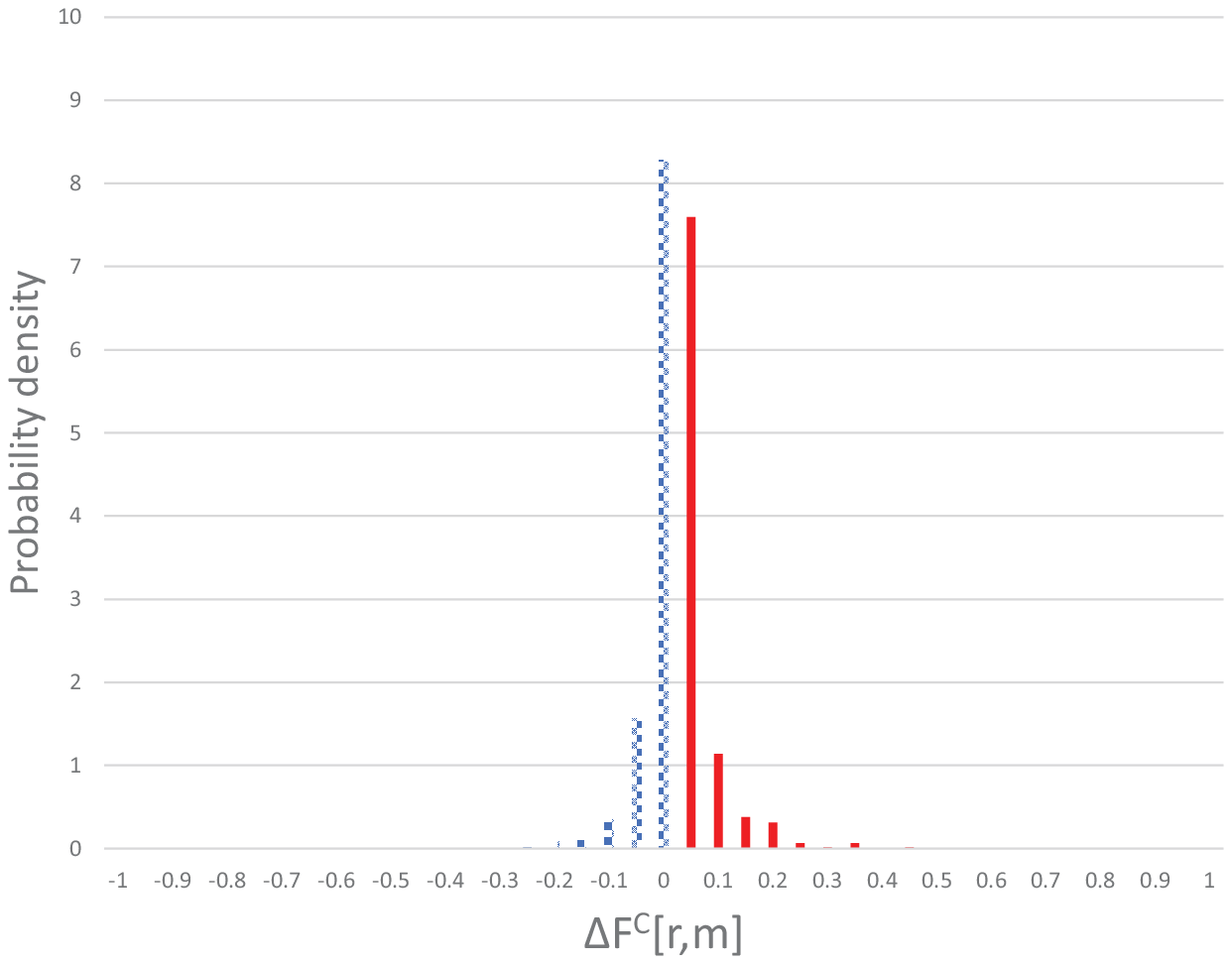}
\put(-300,220){(b)}

\end{center}

%\caption{}

%\label{}

\end{minipage}

\end{tabular}

\caption{(a) Probability density of the correlation coefficients between revenue and market capitalization $C^c{[r,m]}$. (b) Probability density of the difference in VC correlations between revenue and market capitalization $\Delta F^c{[r,m]}$. In (b), we cannot observe asymmetry with respect to zero, which implies that there is no directionality between revenue and market capitalization. See the third line in Table \ref{tab: VC correlation}. In both (a) and (b), the probability density is computed by normalizing the distribution.}
\label{fig:distribution symetry}

\end{figure}

As complementary information, Fig. \ref{fig:distribution assymetry}(a) shows the distribution of the correlation coefficients $C^c{[i,m]}$ between net income(i) and market capitalization(m). Fig. \ref{fig:distribution assymetry}(b) shows the distribution of the difference of VC correlations $F^c{[i,m]}$. Meanwhile, Fig. \ref{fig:distribution symetry}(a) shows the distribution of the correlation coefficient $C^c{[r,m]}$ between revenue(r) and market capitalization(m). Fig. \ref{fig:distribution symetry}(b) shows the distribution of the difference of VC correlation $F^c{[r,m]}$.

We can observe that the distribution of the difference in VC correlations $F^c{[i,m]}$ in Fig. \ref{fig:distribution assymetry}(b) is asymmetric compared to Fig. \ref{fig:distribution symetry}(b). This implies that there is directionality from net income to market capitalization. Although the asymmetry appears to be small in Fig. \ref{fig:distribution assymetry}(b), the large sample size makes the p-value very small, as shown in the first row of Table \ref{tab: VC correlation} (p-value is $6.49\times10^{-9}$). This indicates that there is sufficient statistical significance for the suggested directionality.

As shown in Table \ref{tab: correlation coefficients} and Fig. \ref{fig:Correlation network}, all the correlations between the accounting variables are positively correlated. However, some pairs of accounting variables have not only positive correlations but also directionality of influence with statistical significance, as shown in Table \ref{tab: VC correlation}. From the directionality of the VC correlations presented in Table \ref{tab: VC correlation}, we show the directionality network in Fig. \ref{fig:Directionality}. The five pairs of accounting variables have directionality of influence with statistical significance, and we show them as five oriented links in Fig. \ref{fig:Directionality}. A comparison of Fig. \ref{fig:Directionality} with Fig. \ref{fig:Correlation network} clearly shows the directionality between the accounting variables.

\subsection{Statistical validation analysis}
To further validate these results, we perform two statistical tests. The first validation requires splitting the data into two groups and validating the results for each group independently. For the second test (randomization), we reshuffle the data to check whether the results can be verified.

For the first statistical test, we split the data of all companies into two groups of equal numbers (each part has half). Specifically, we arrange the data in order of decreasing security codes and divide them into an odd-numbered group and an even-numbered group. We then compute the VC correlation for these two groups in the same way as for the full dataset. 

The statistical analysis results are presented in Table \ref{tab: VC correlation validation splitting}. All the results with statistical significance in the splitting data are consistent with the directionality result of the original non-splitting result. Note that when the sample size decreases, the statistical significance tends to decrease. However, the test results indicate that the directionality signal was sufficiently robust and consistent with the original results (directionality) with a raised threshold of 0.2. Therefore, we conclude that the splitting data analysis validated the original directionality.

Second, we randomize the data by randomly swapping the accounting variables for each company to calculate the VC correlation. We show the results in Table \ref{tab: VC correlation validation randomized}. We compute all pairs whose directionality is confirmed in the original data. As expected, the statistical significance of directionality completely disappears for all pairs. This means that even when the threshold is raised to 0.2, the statistical signal of directionality vanishes, as expected. This indicates that the directional statistical signal of the original calculation is reliable.

In summary, we have validated the results using two methods, which significantly support the robustness and reliability of the original predictions.

\subsection{Determination of threshold $h$}
\label{sec:threshold}

The present analysis has been done with the threshold $h$ set up as 0.2. In what follows, we will discuss how the results change when we vary the threshold $h$.

\begin{figure}[htbp]
%\vspace{-8cm}
\begin{tabular}{cc}
%\vspace{-4cm}
\begin{minipage}{0.5\hsize}
\begin{center}
\includegraphics[scale=0.25]{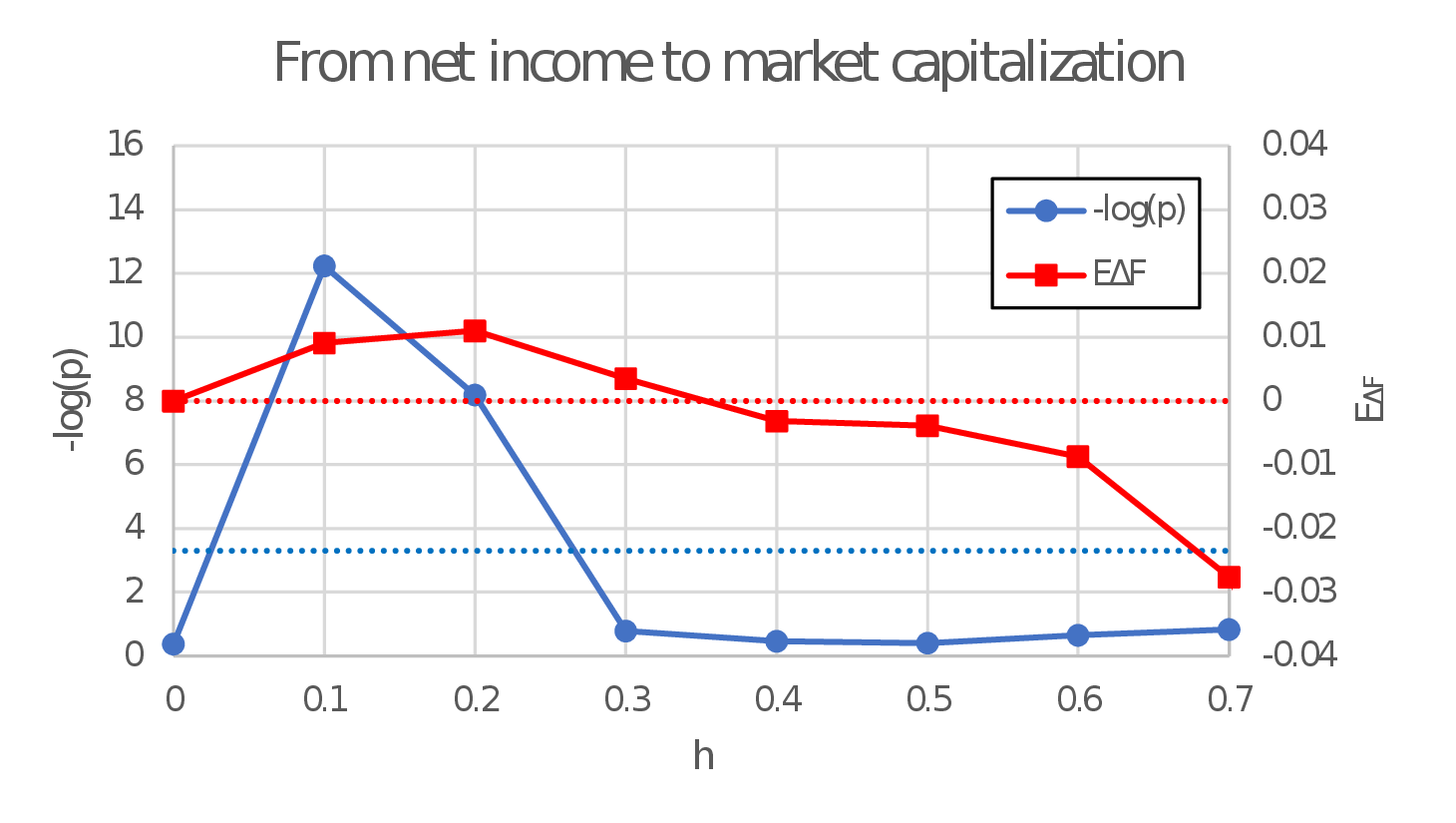}
\put(-180,90){(a)}
\end{center} 
\end{minipage} 
\begin{minipage}{0.5\hsize}
\begin{center}
\includegraphics[scale=0.25]{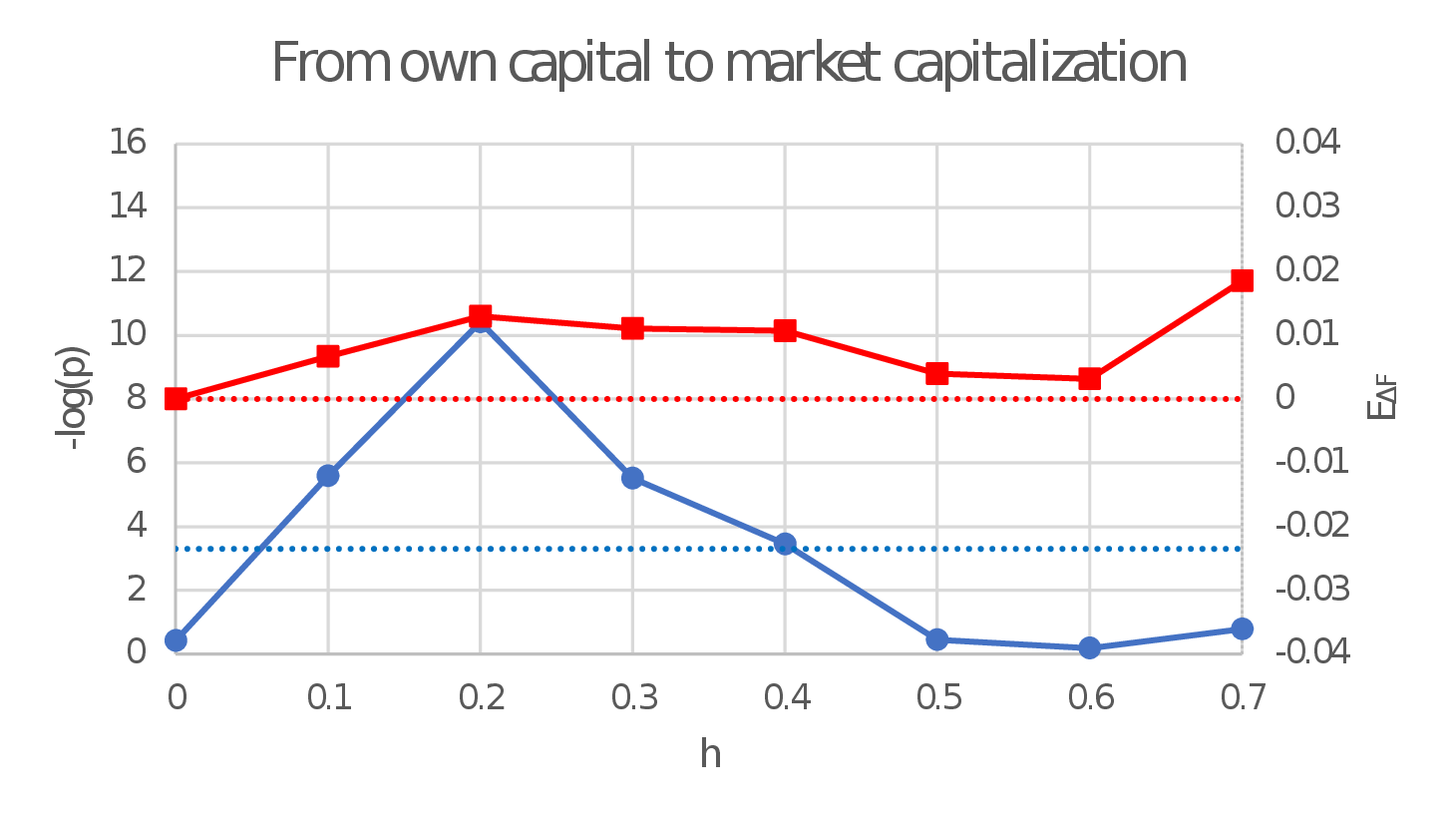}
\put(-180,90){(b)}
\end{center}
\end{minipage}\\

%\vspace{-4cm}
\begin{minipage}{0.5\hsize}
\begin{center}
\includegraphics[scale=0.25]{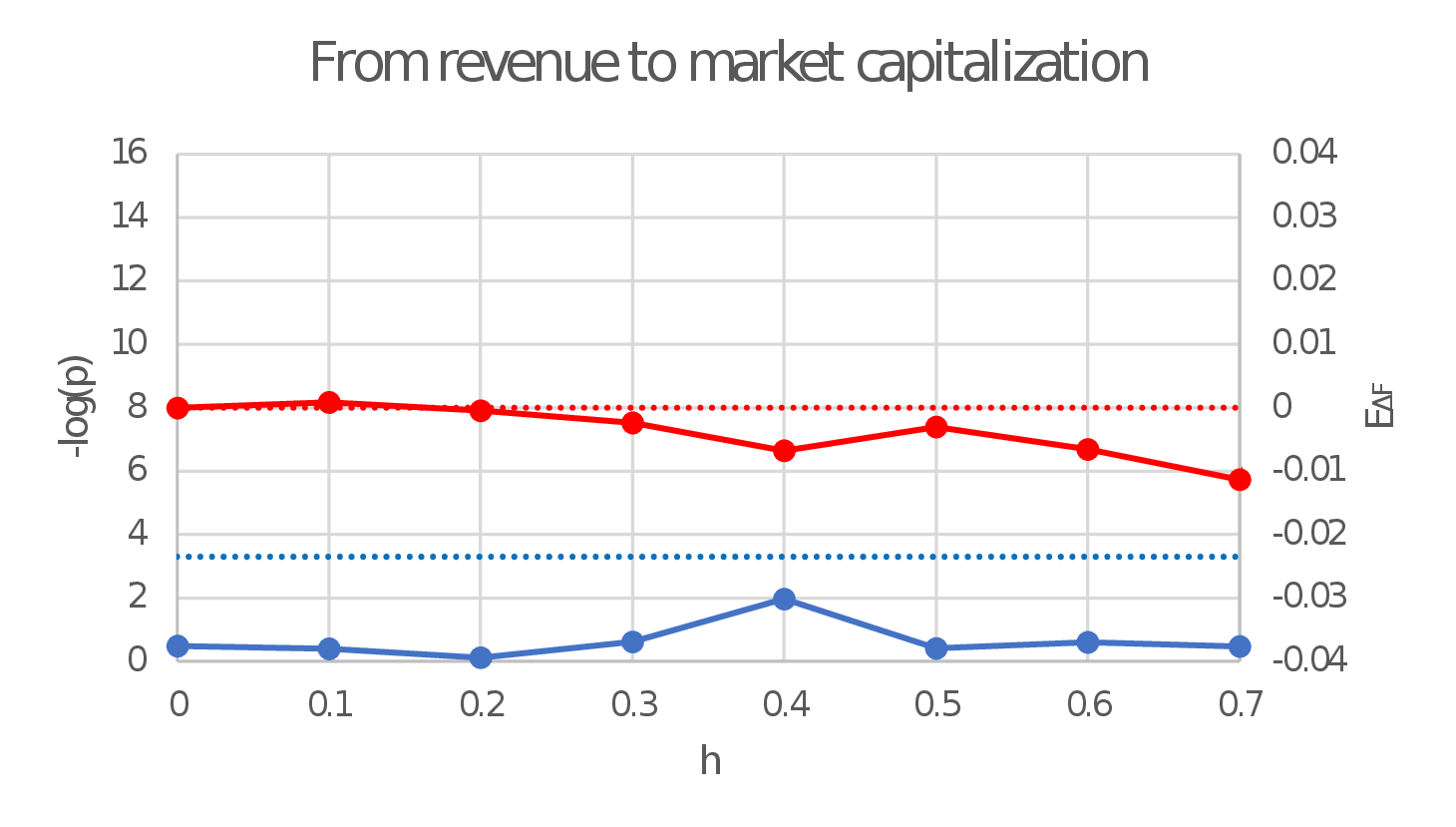}
\put(-180,90){(c)}
\end{center}
\end{minipage}
\begin{minipage}{0.5\hsize}
\begin{center}
\includegraphics[scale=0.25]{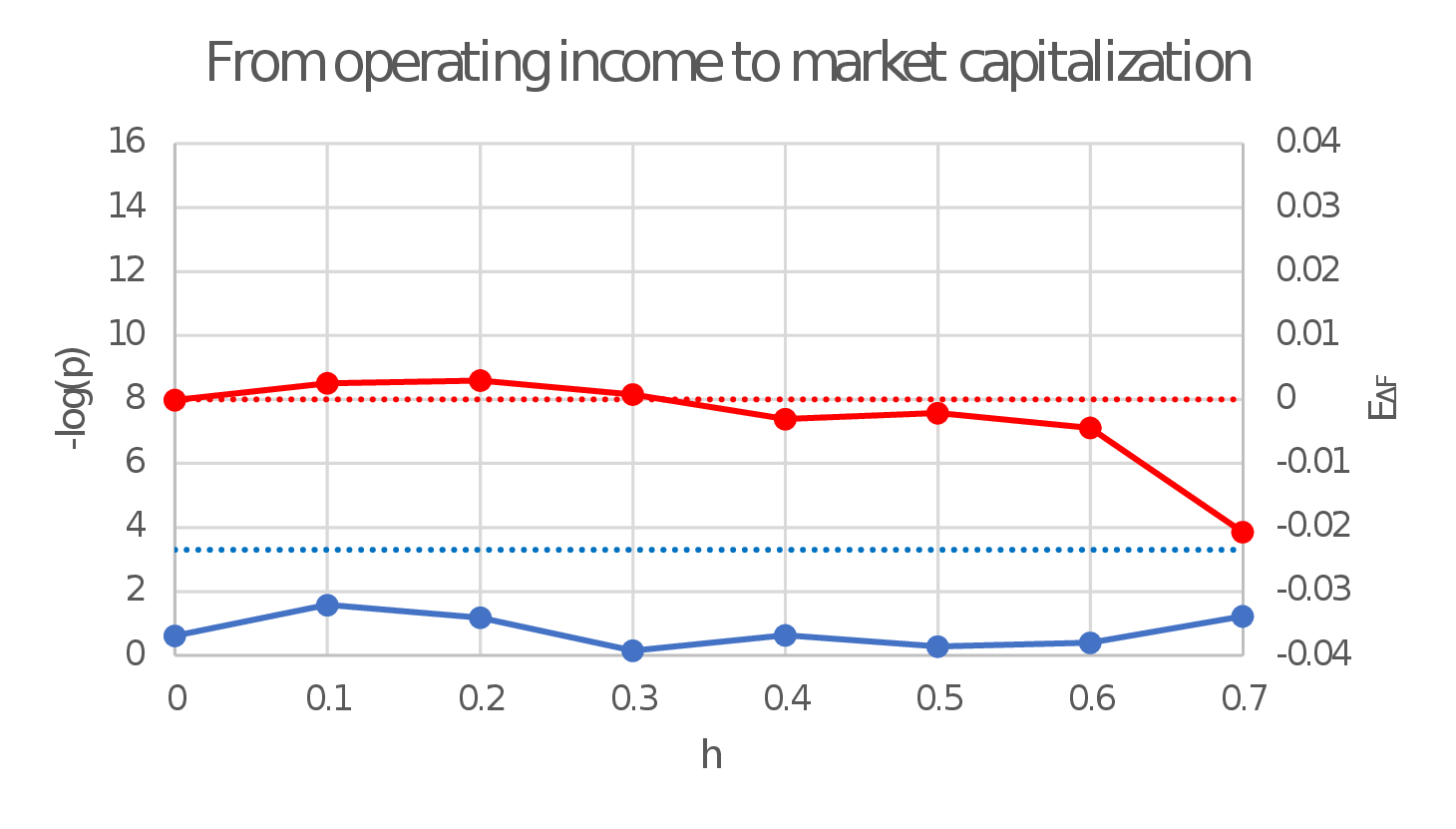}
\put(-180,90){(d)}
\end{center}
\end{minipage}\\

%\vspace{-4cm}
\begin{minipage}{0.5\hsize}
\begin{center}
\includegraphics[scale=0.25]{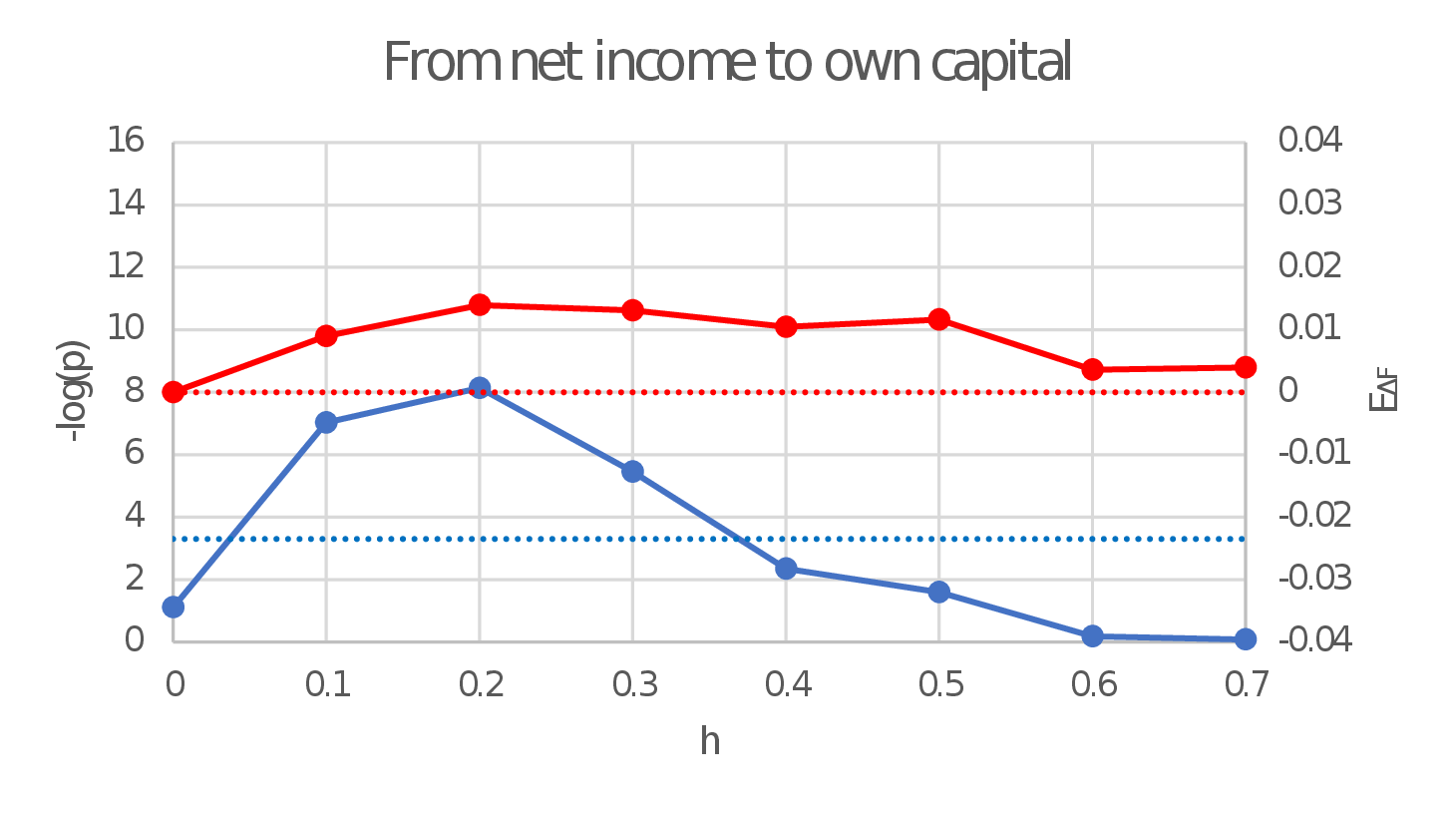}
\put(-180,90){(e)}
\end{center}
\end{minipage}
\begin{minipage}{0.5\hsize}
\begin{center}
\includegraphics[scale=0.25]{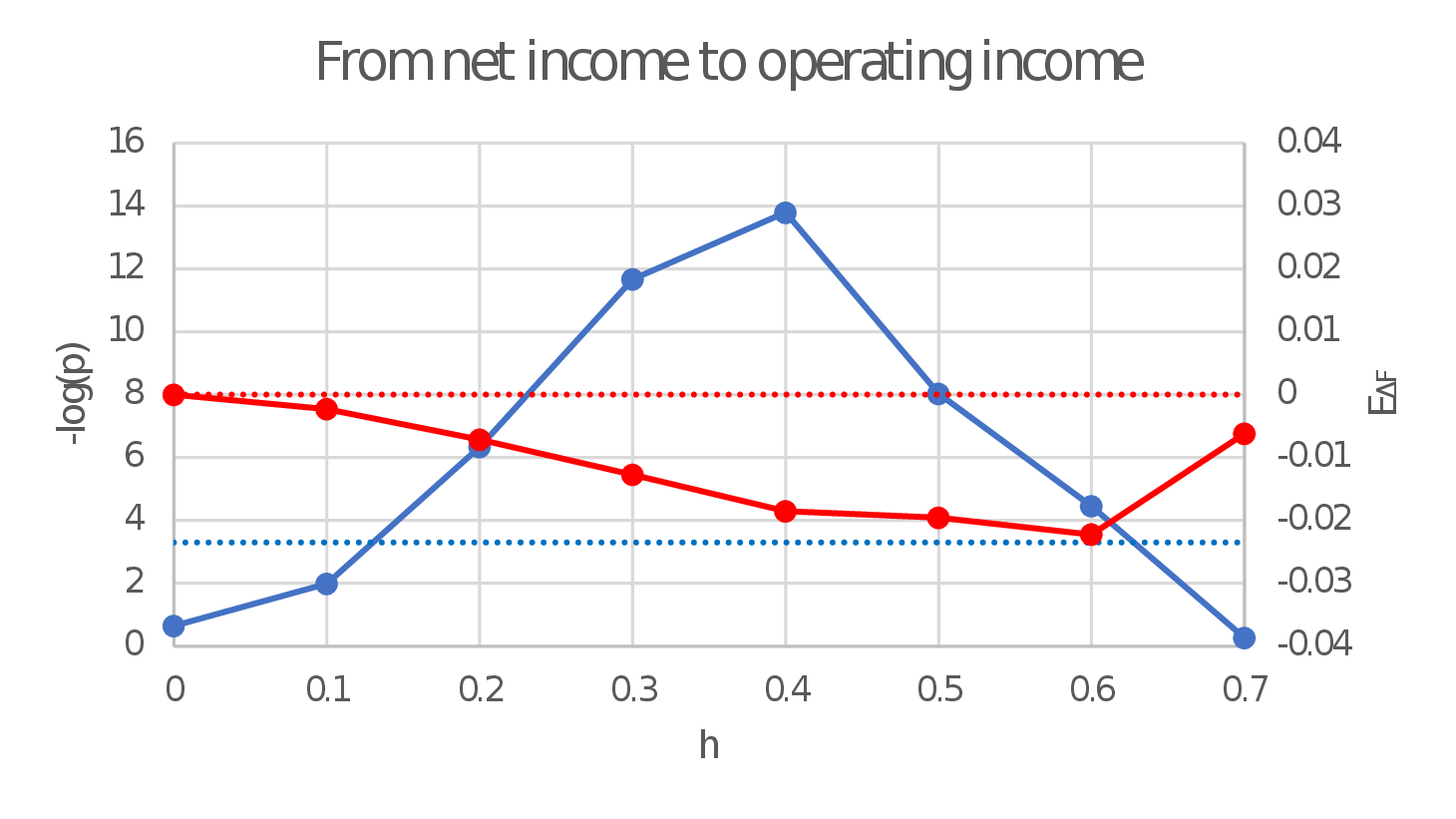}
\put(-180,90){(f)}
\end{center}
\end{minipage}\\

%\vspace{-4cm}
\begin{minipage}{0.5\hsize}
\begin{center}
\includegraphics[scale=0.25]{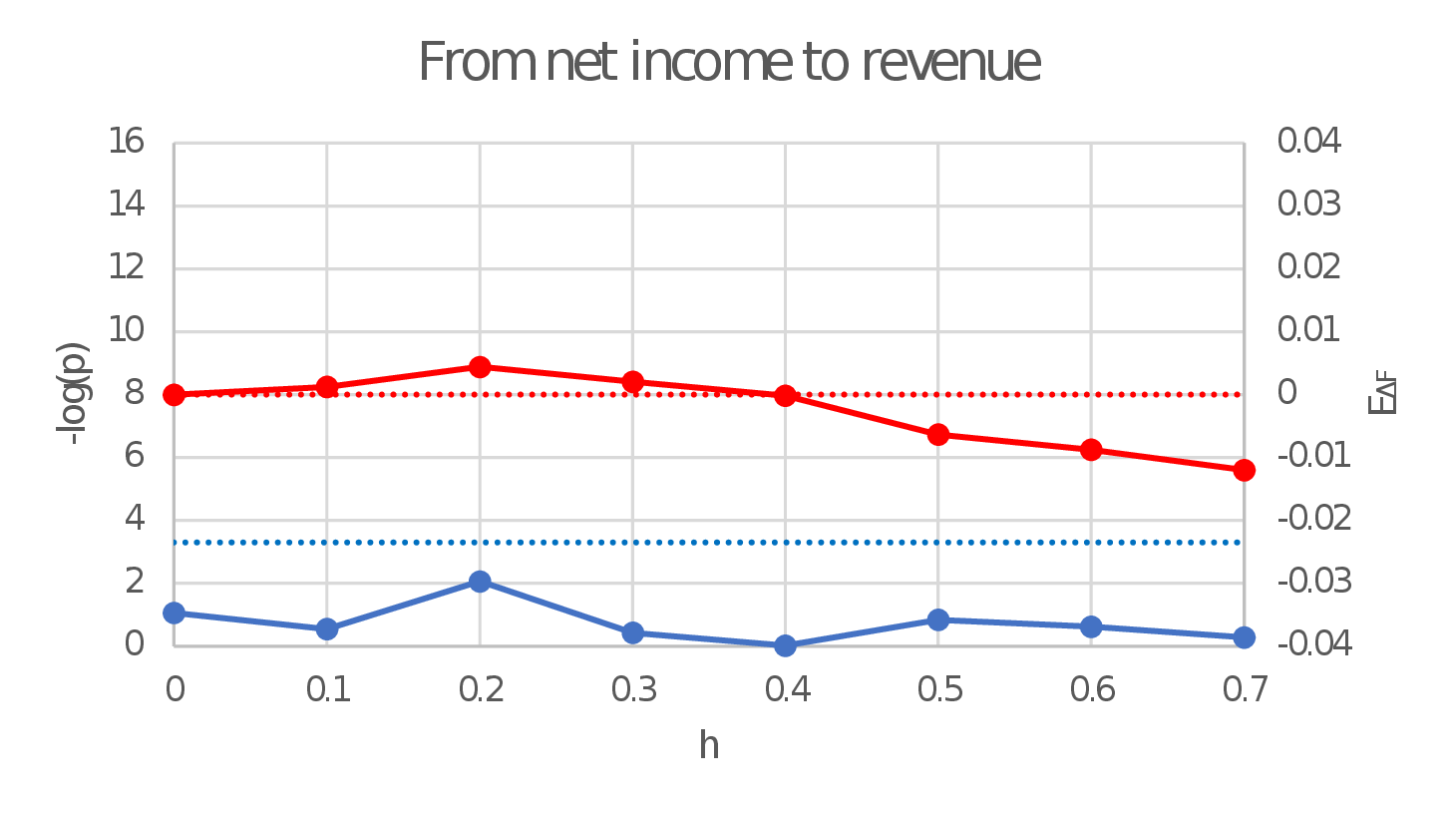}
\put(-180,90){(g)}
\end{center}
\end{minipage}
\begin{minipage}{0.5\hsize}
\begin{center}
\includegraphics[scale=0.25]{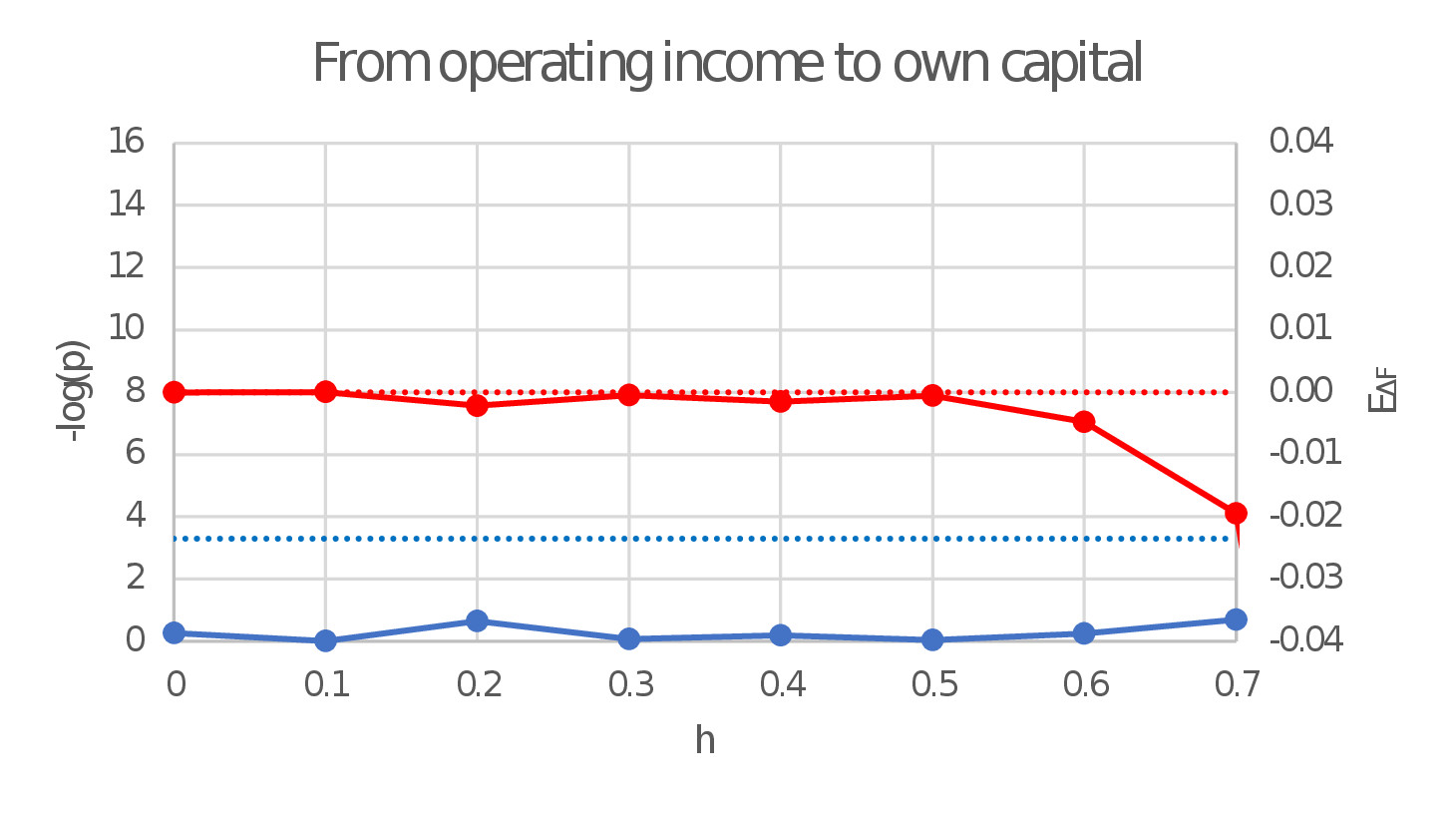}
\put(-180,90){(h)}
\end{center}
\end{minipage}\\

%\vspace{-4cm}
\begin{minipage}{0.5\hsize}
\begin{center}
\includegraphics[scale=0.25]{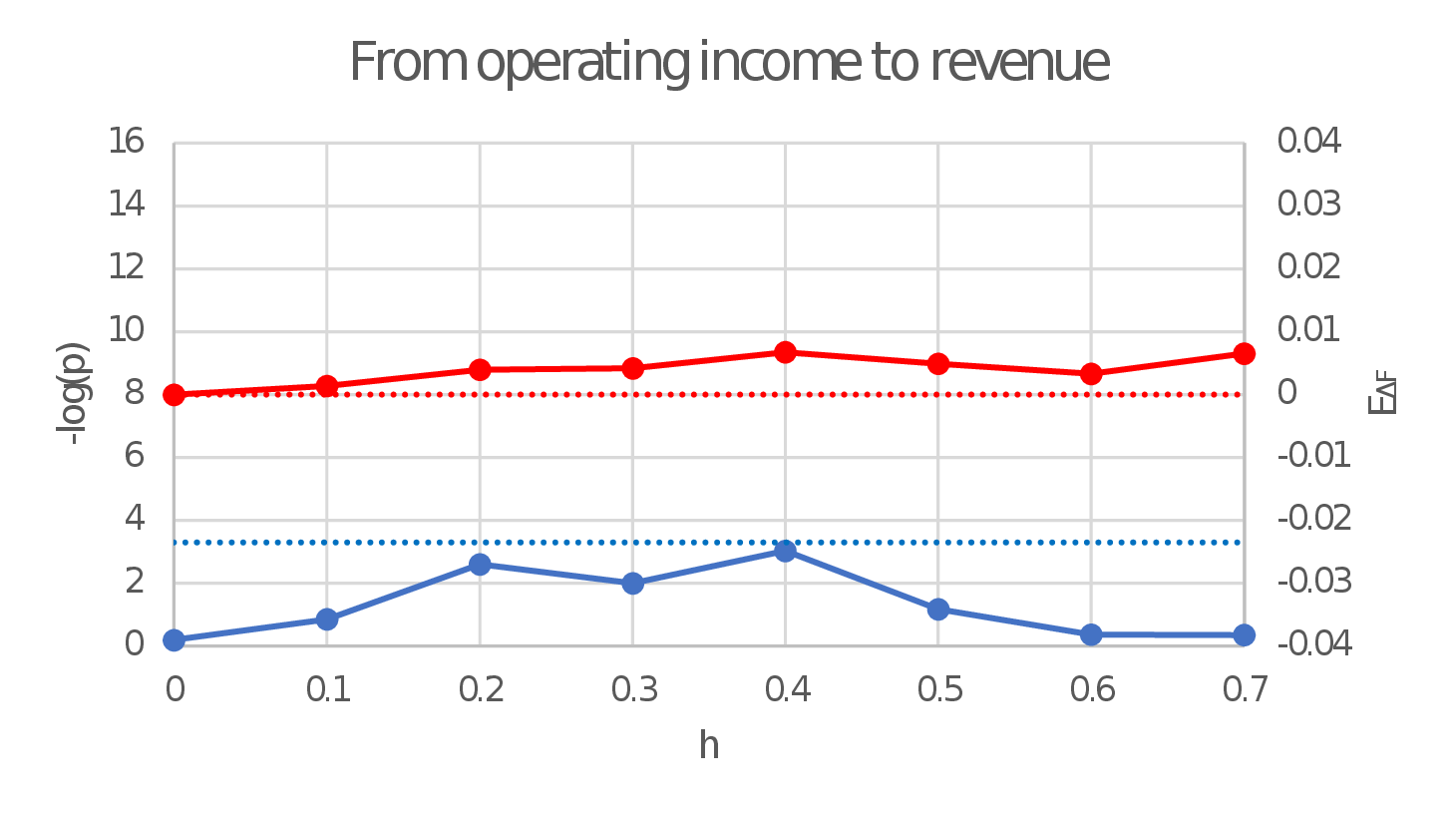}
\put(-180,90){(i)}
\end{center}
\end{minipage}
\begin{minipage}{0.5\hsize}
\begin{center}
\includegraphics[scale=0.25]{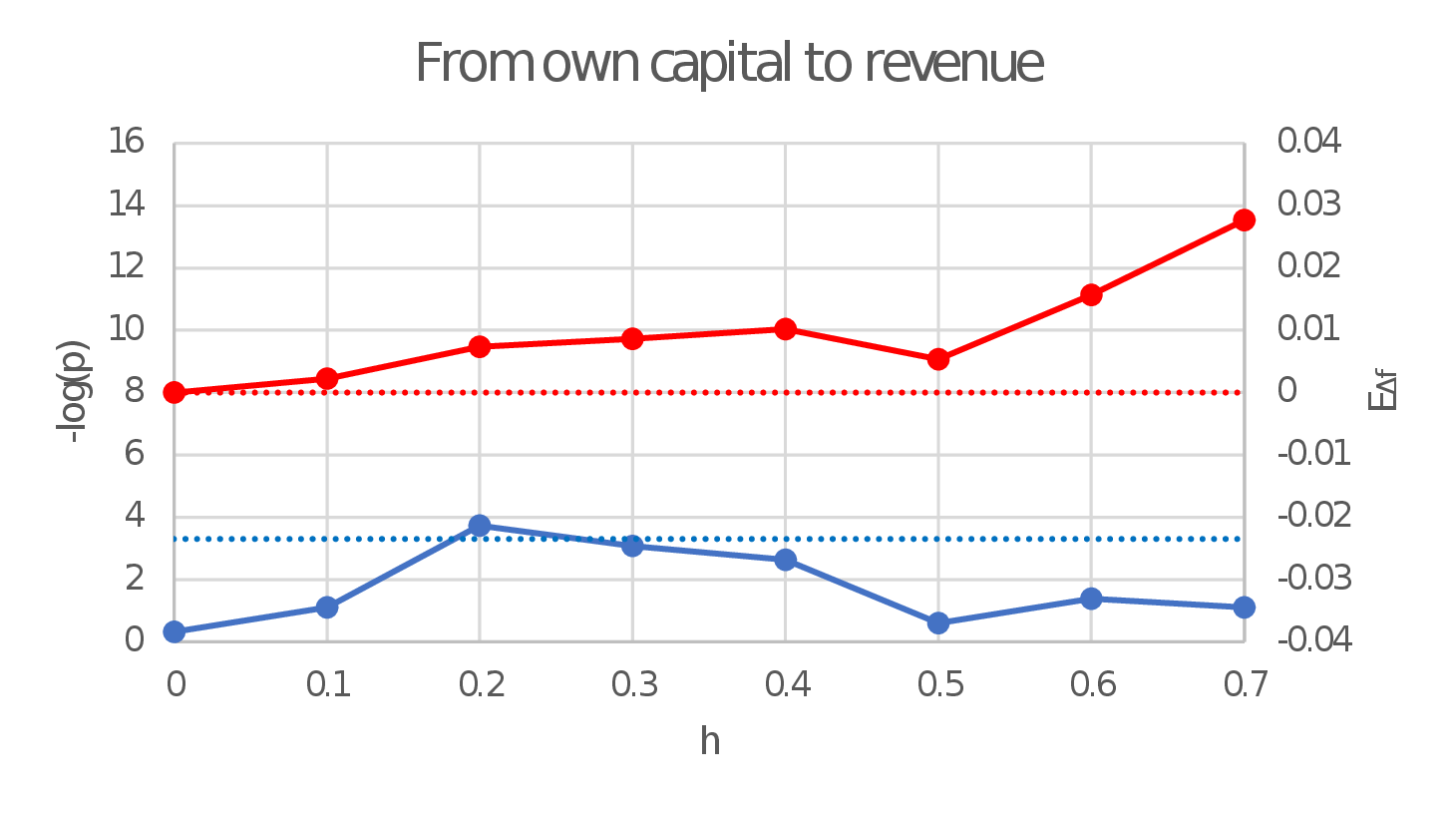}
\put(-180,90){(j)}
\end{center}
\end{minipage}

\end{tabular}

%\vspace{4cm}
\caption{$E_{\Delta F}$ and p-value are displayed as a function of the threshold value $h$. The horizontal axis denotes the threshold $h$ and the vertical axis indicates $E_{\Delta F}$ and $-\log(p)$. The blue dotted line indicates $p=p_0$, and the red dotted line indicates $E_{\Delta F}=0$.}

\label{fig:directionality as a function of threshold}

\end{figure}

In Fig. \ref{fig:directionality as a function of threshold}, for each accounting variable pair $[s,s^\prime]$, we have displayed $E_{\Delta F}{[s,s^\prime]}$ and p-value as a function of the threshold $h$. (That is, the horizontal axis is the threshold $h$, and the vertical axis is $E_{\Delta F}$ for directionality and $-\log(p)$ for statistical significance.) The blue dotted line indicates the threshold value of $p=p_0$. The red dotted line indicates $E_{\Delta F}=0$. If the blue line exceeds the blue dotted line, there is a statistical significance. Furthermore, in that case, if $E_{\Delta F}$ (red line) is positive (above the red dotted line), then a directionality is predicted from $s$ to $s^\prime$. Conversely, if it is negative, a directionality is predicted from $s^\prime$ to $s$.

The relation between the statistical significance of directionality predictions and the effect of the threshold parameter $h$ is shown in Fig. \ref{fig:directionality as a function of threshold}. As a first example, in Fig. \ref{fig:directionality as a function of threshold}(a), when $h=0.2$ and $0.3$, there is a statistical significance (blue line is above the blue dotted line) and $E_{\Delta F}$ (red line) is positive. This indicates that there is a directionality from net income to market capitalization. Next, in the case of Fig. \ref{fig:directionality as a function of threshold}(f), there is statistical significance when $h = 0.2, 0.3, 0.4, 0.5$ and $0.6$. In that case, $E_{\Delta F}$ is negative, so the direction is from operating income to net income. In the case of Fig. \ref{fig:directionality as a function of threshold}(c), the blue line never exceeds the blue dotted line, so there is no statistical significance for any threshold $h$. Therefore, there is no directional relationship between revenue and market capitalization. All results in Fig. \ref{fig:directionality as a function of threshold} show that the directionality does not change when the threshold $h$ is shifted to some extent, indicating that the results are robust to these changes.

Fig. \ref{fig:threshold and directional network} shows how the structure of the directional network changes when the threshold $h$ is varied. The thickness of the oriented links is proportional to $-\log(p)$, and the greater the thickness, the greater the statistical significance. As shown in Fig. \ref{fig:threshold and directional network}, as the threshold $h$ increases, the number of links increases, reaching the maximum number of links and link thickness at $h=0.2$. Moreover, when $h$ increases further, all the links vanish progressively. Since we have the number of links as well as the link strength (thickness) reach a maximum when the threshold $h$ is 0.2, we selected the threshold $h=0.2$ for the computation as shown in Section \ref{sec:VC correlation analysis predictions}. It is worth noting that as the threshold $h$ increases, the directional network structure is progressively unveiled starting  from zero number of links ($h=0$), and reaches its maximum at $h=0.2$. From this point, the links start vanishing, driven by a decreasing of the statistical significance ($h=0.7$).

\subsection{Interpretation of the results}

The accounting variables in our analysis belong to several categories of financial accounting. Revenue, operating income, and net income are the main accounting variables in the income statement. In accounting terms, revenue is calculated first, then operating income, and finally, net income. Own capital is an accounting variable found in the balance sheet and is equivalent to total assets minus liabilities. Market capitalization (or market cap) is the total market value of publicly traded shares and is equivalent to the share price multiplied by the number of shares. By bearing these key concepts in mind, we can derive the following findings from our data-driven analysis.

\begin{figure}[htbp]

%\begin{center}

\hspace*{-4cm}
\includegraphics[scale=0.6]{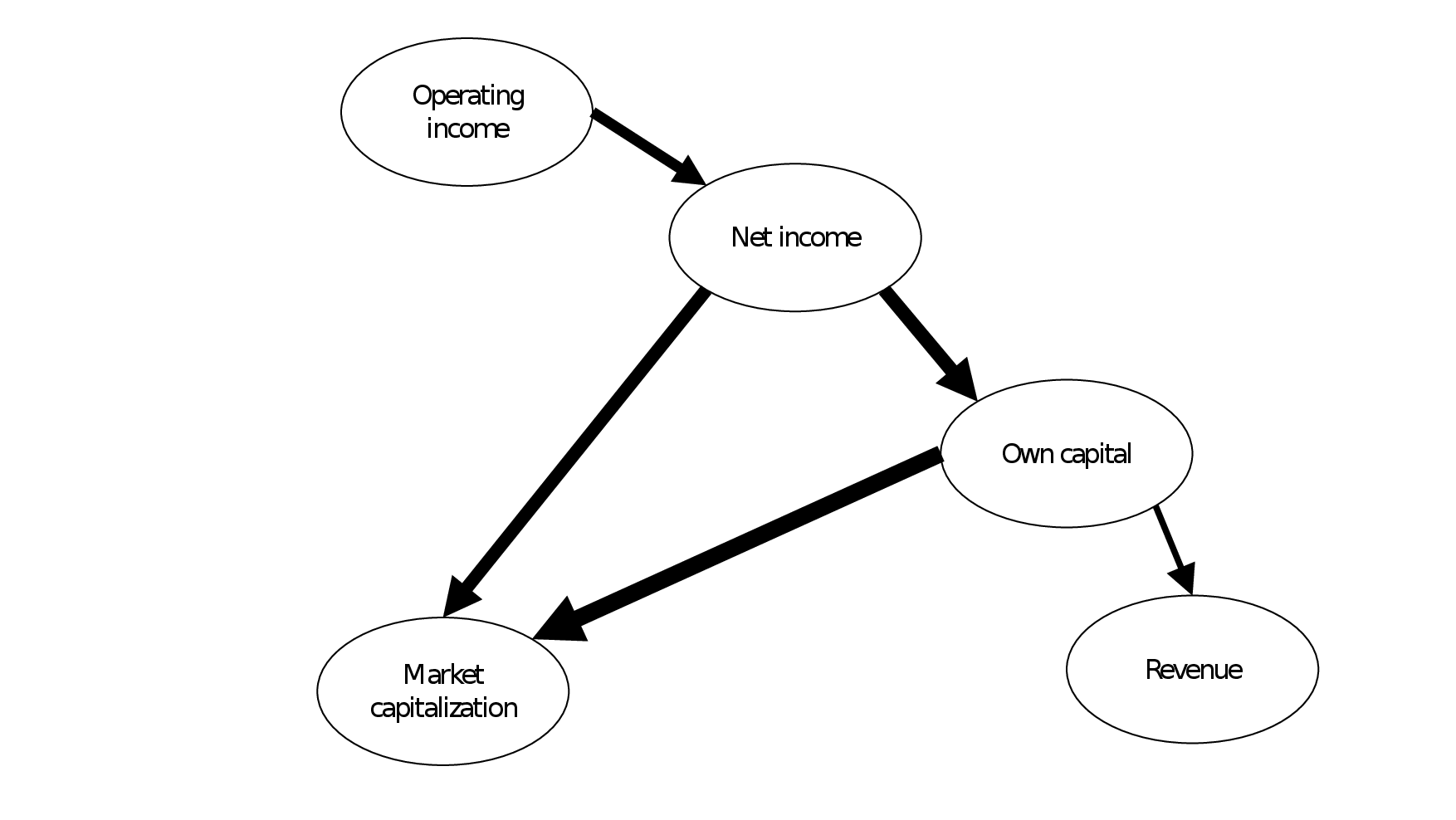}
%\includegraphics[scale=0.6]{directionality-p-value-3.eps}

%\put(-200,150){(b)}

%\end{center}

\caption{Directionality of the five major accounting variables for the case of threshold $h=0.2$. Thickness denotes the statistical significance of directionality, since it is linearly determined by the minus of the logarithm of the p-value($p$) (i.e. $-\log(p)$) in Table \ref{tab: VC correlation}. Note that in Fig. \ref{fig:Correlation network}, directionality is not predicted as in Fig \ref{fig:Directionality}.}

\label{fig:Directionality}

\end{figure}

\begin{figure}[htbp]
%\vspace{-8cm}
\begin{tabular}{cc}
%\vspace{-4cm}
\begin{minipage}{0.5\hsize}
\begin{center}
\fbox{
\includegraphics[scale=0.2]{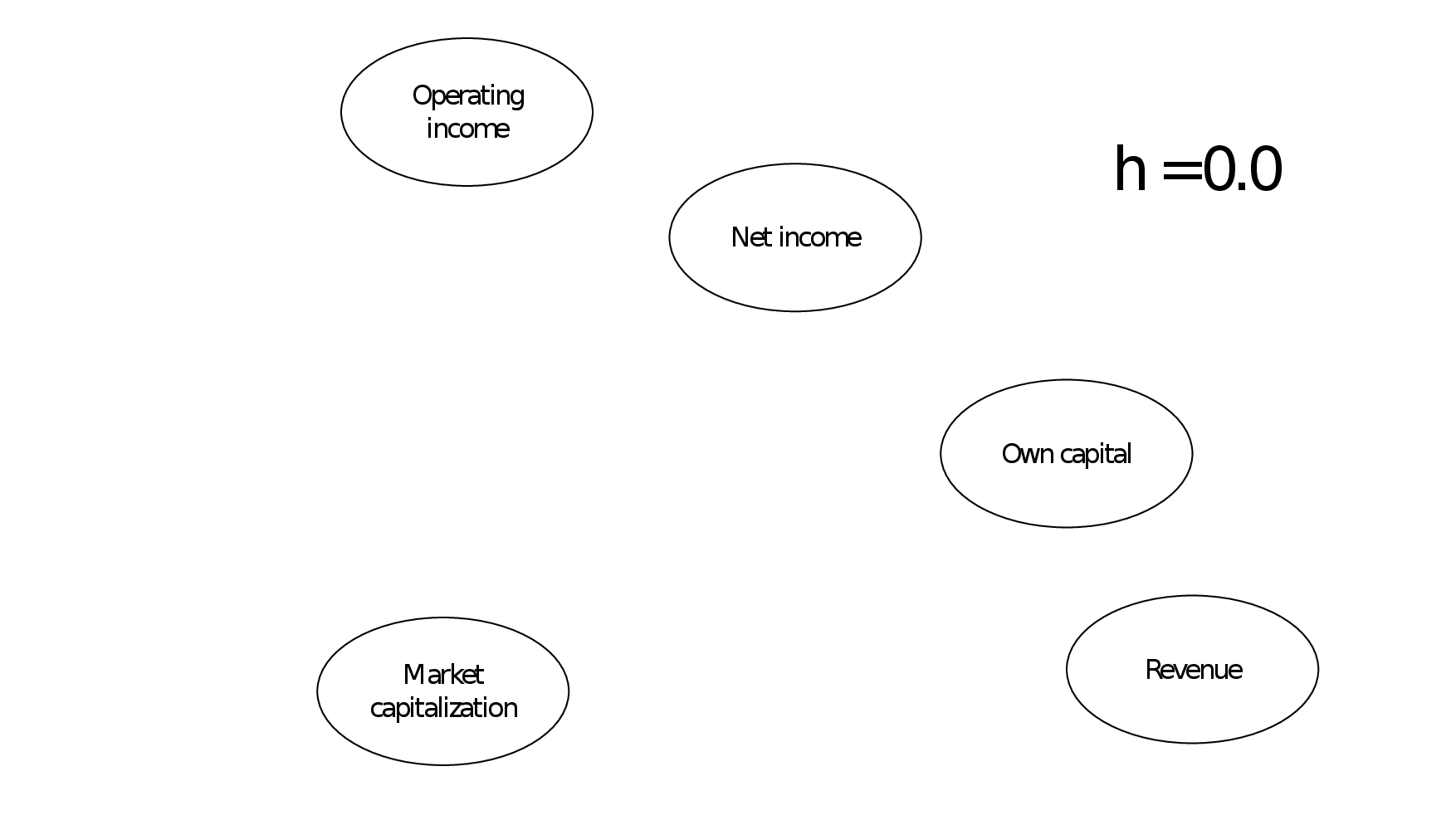}
}
\put(-170,75){(1)}
\end{center} 
\end{minipage} 
\begin{minipage}{0.5\hsize}
\begin{center}
\fbox{
\includegraphics[scale=0.2]{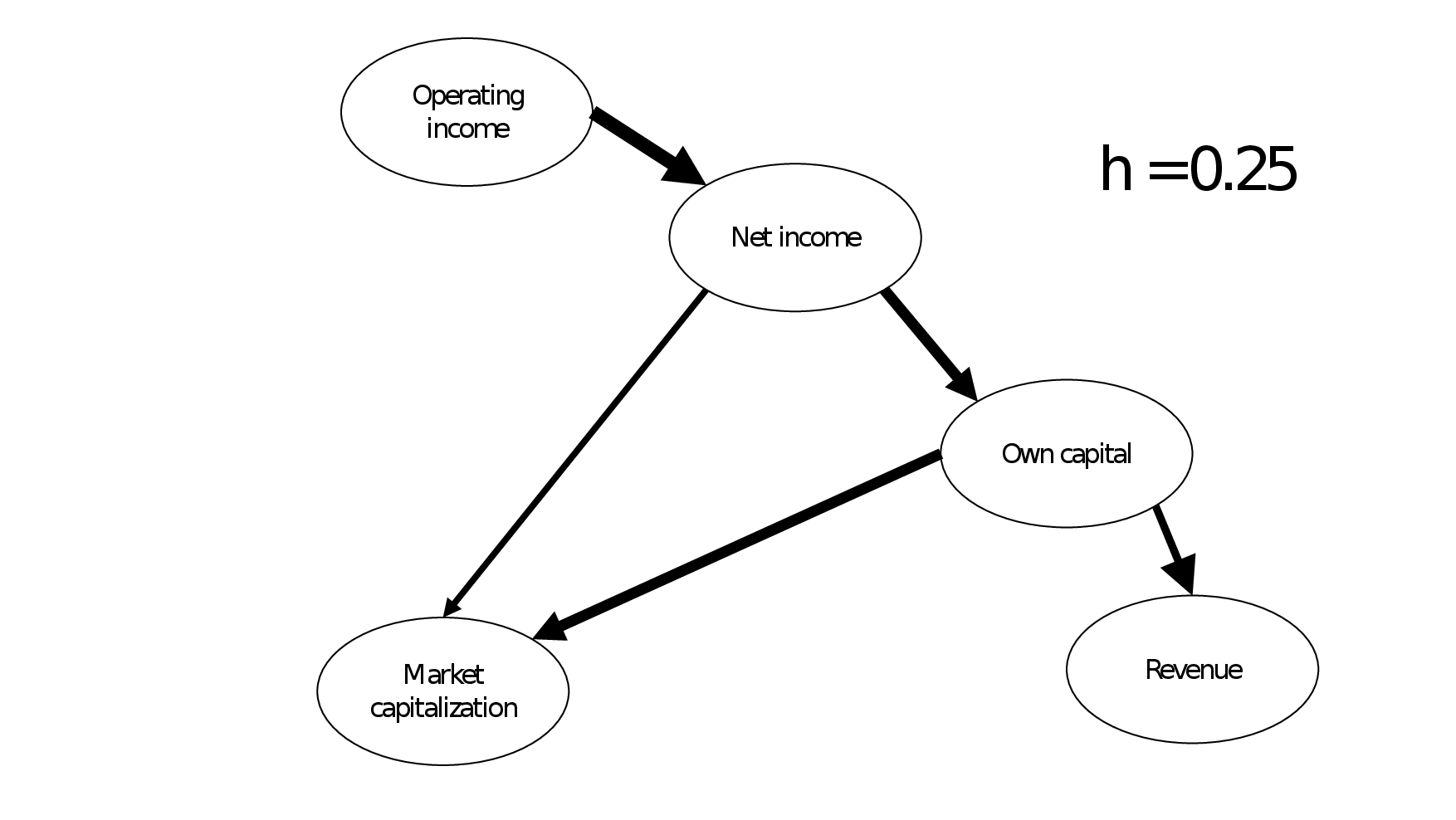}
}
\put(-170,75){(6)}
\end{center}
\end{minipage}\\

%\vspace{-4cm}
\begin{minipage}{0.5\hsize}
\begin{center}
\fbox{
\includegraphics[scale=0.2]{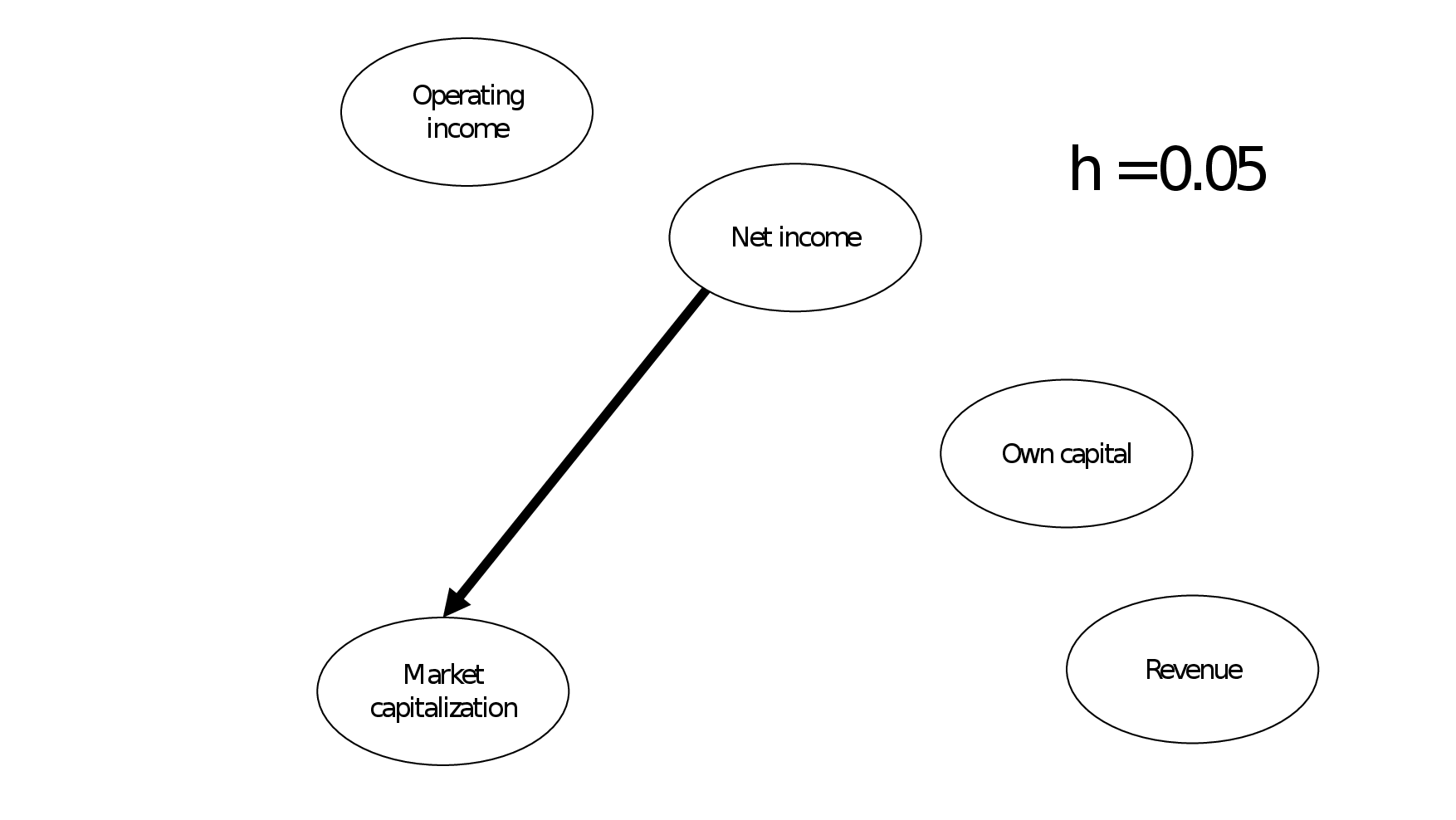}
}
\put(-170,75){(2)}
\end{center}
\end{minipage}
\begin{minipage}{0.5\hsize}
\begin{center}
\fbox{
\includegraphics[scale=0.2]{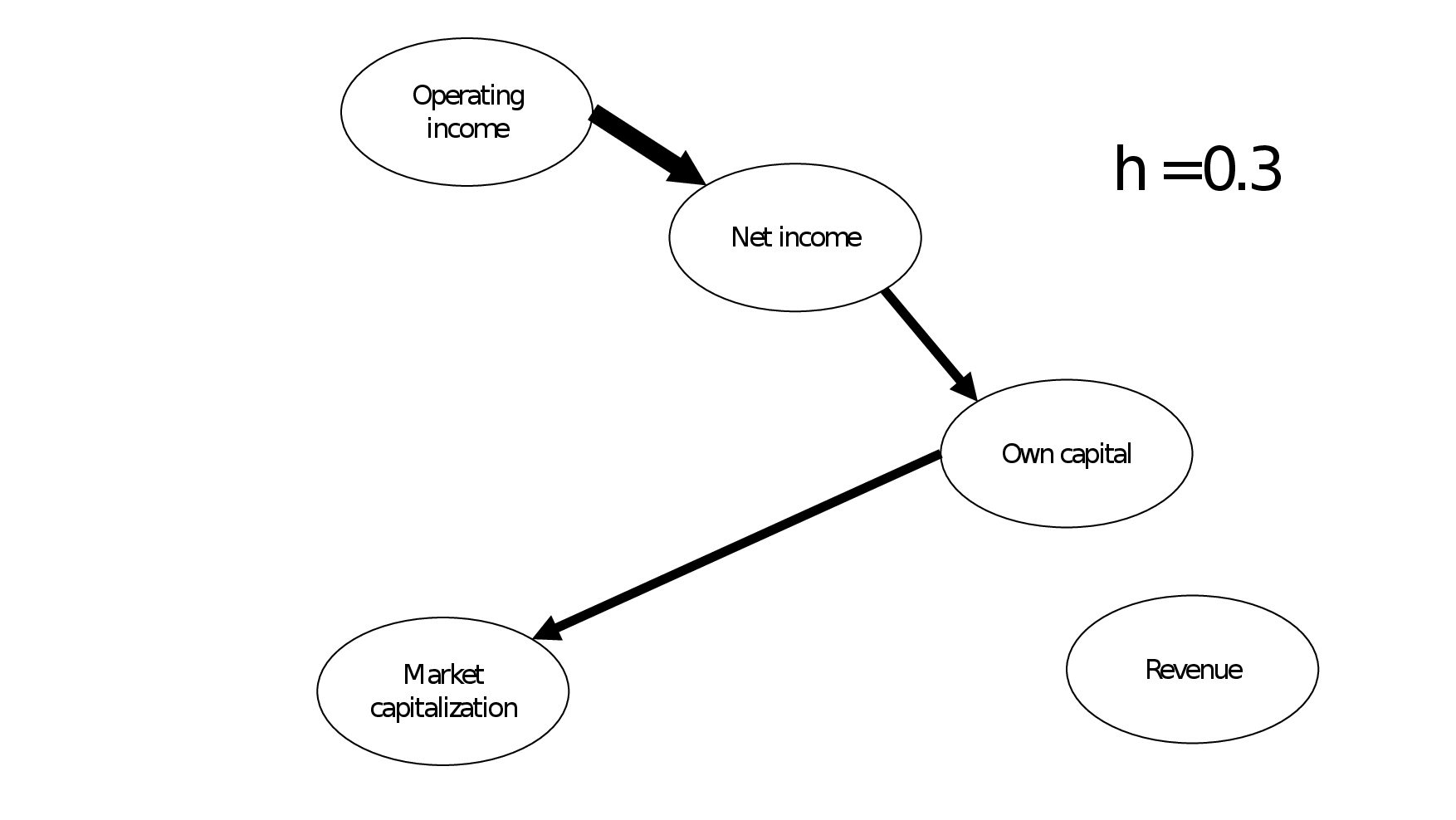}
}
\put(-170,75){(7)}
\end{center}
\end{minipage}\\

%\vspace{-4cm}
\begin{minipage}{0.5\hsize}
\begin{center}
\fbox{
\includegraphics[scale=0.2]{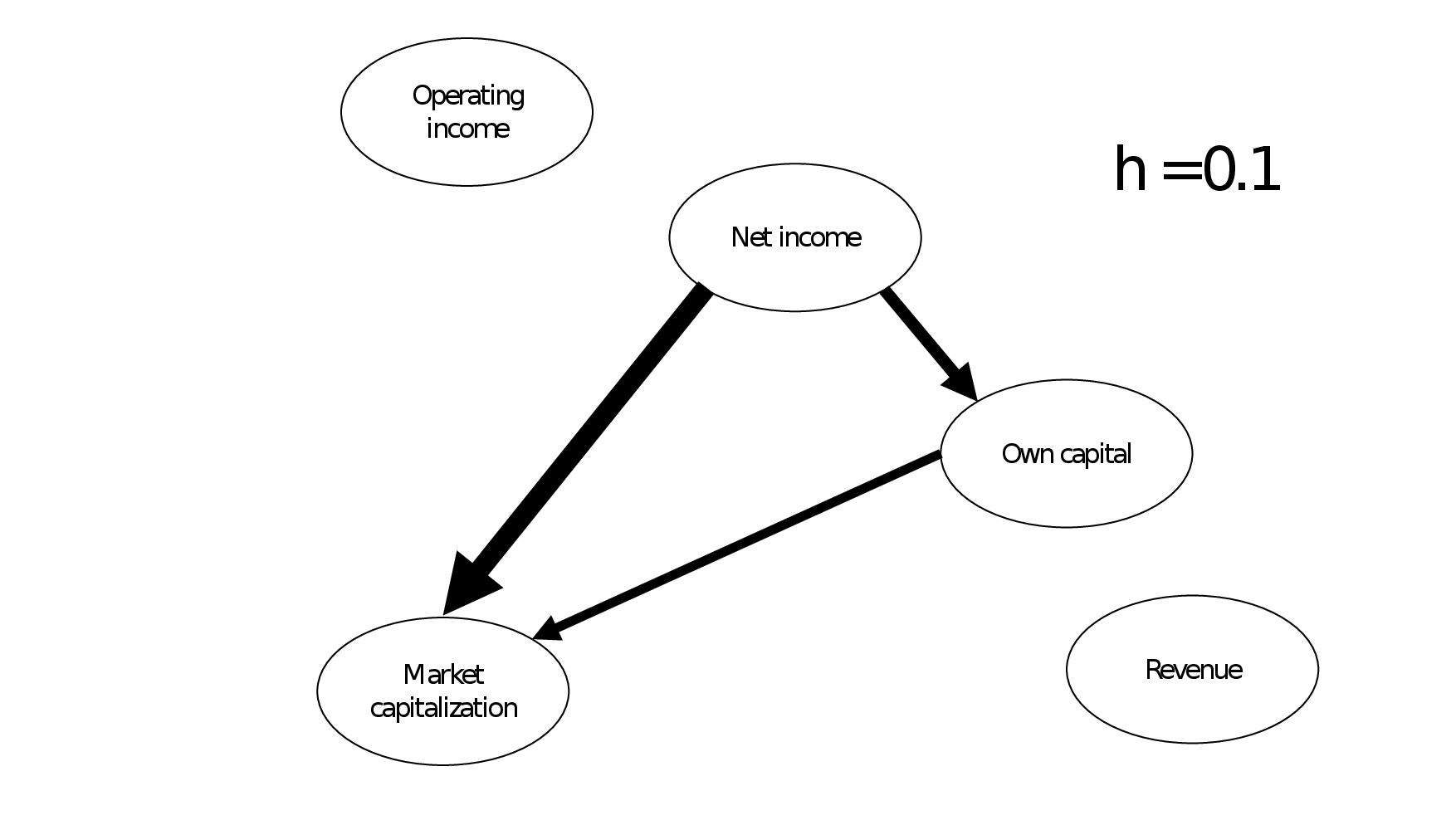}
}
\put(-170,75){(3)}
\end{center}
\end{minipage}
\begin{minipage}{0.5\hsize}
\begin{center}
\fbox{
\includegraphics[scale=0.2]{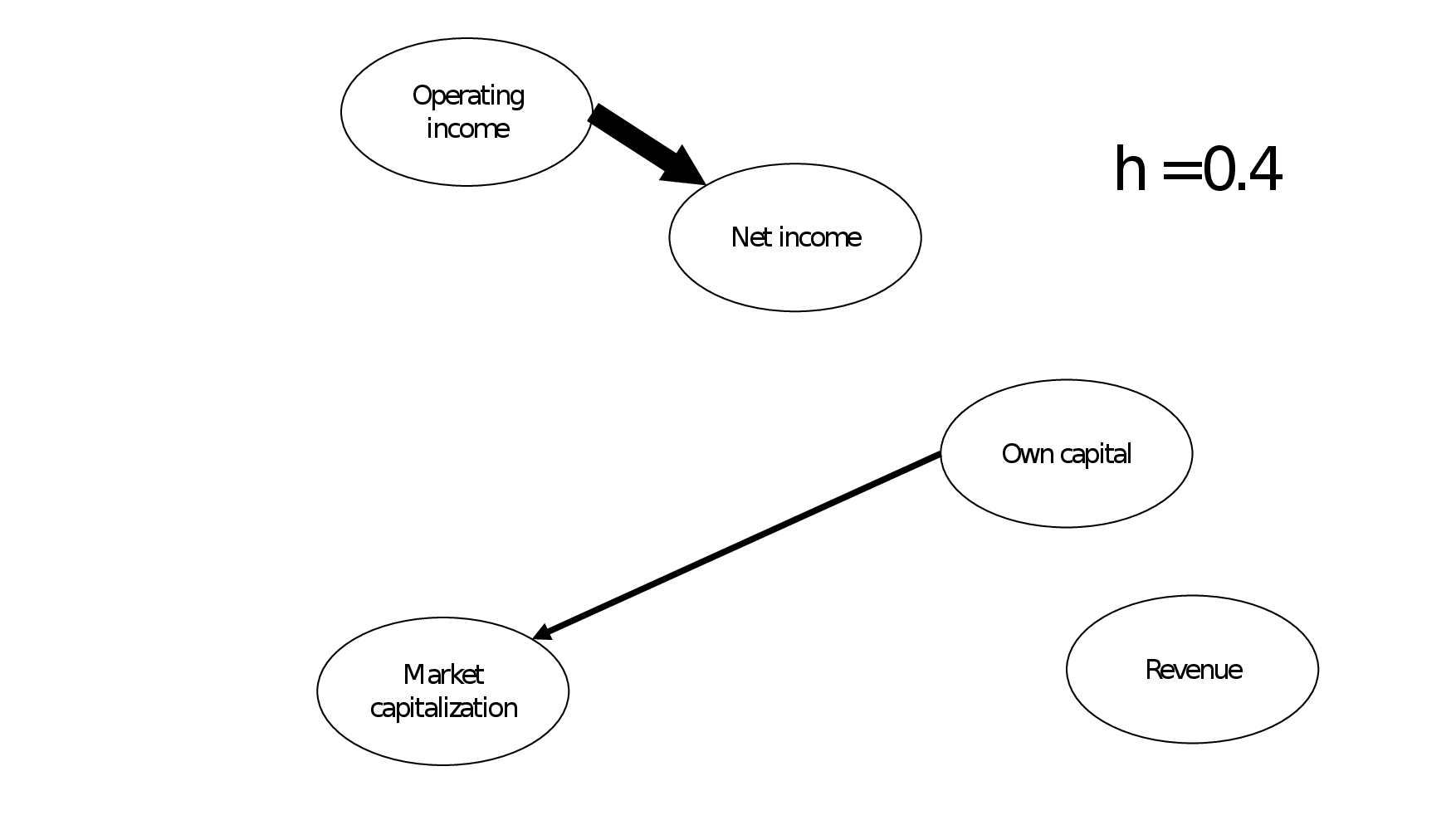}
}
\put(-170,75){(8)}
\end{center}
\end{minipage}\\

%\vspace{-4cm}
\begin{minipage}{0.5\hsize}
\begin{center}
\fbox{
\includegraphics[scale=0.2]{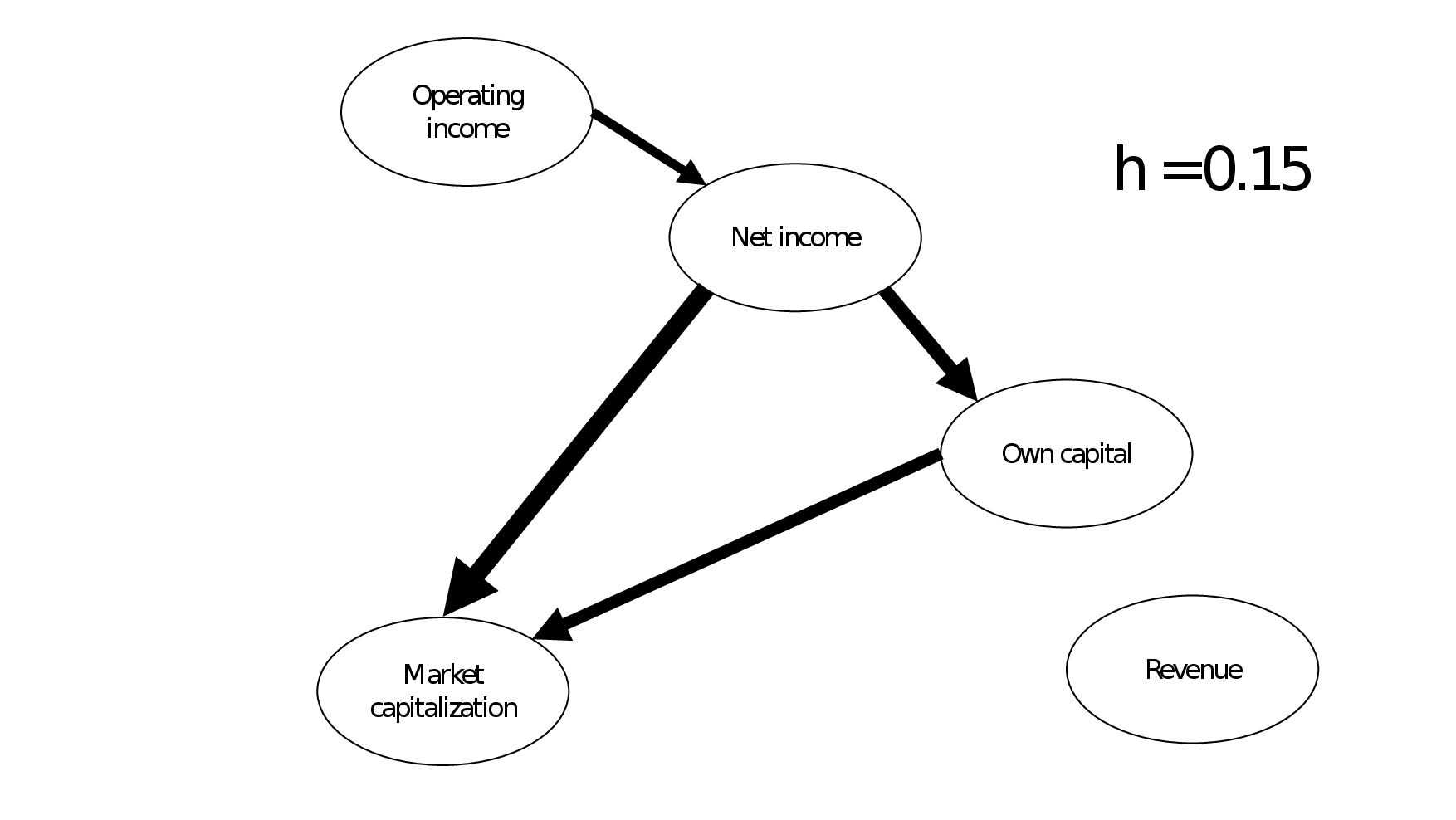}
}
\put(-170,75){(4)}
\end{center}
\end{minipage}
\begin{minipage}{0.5\hsize}
\begin{center}
\fbox{
\includegraphics[scale=0.2]{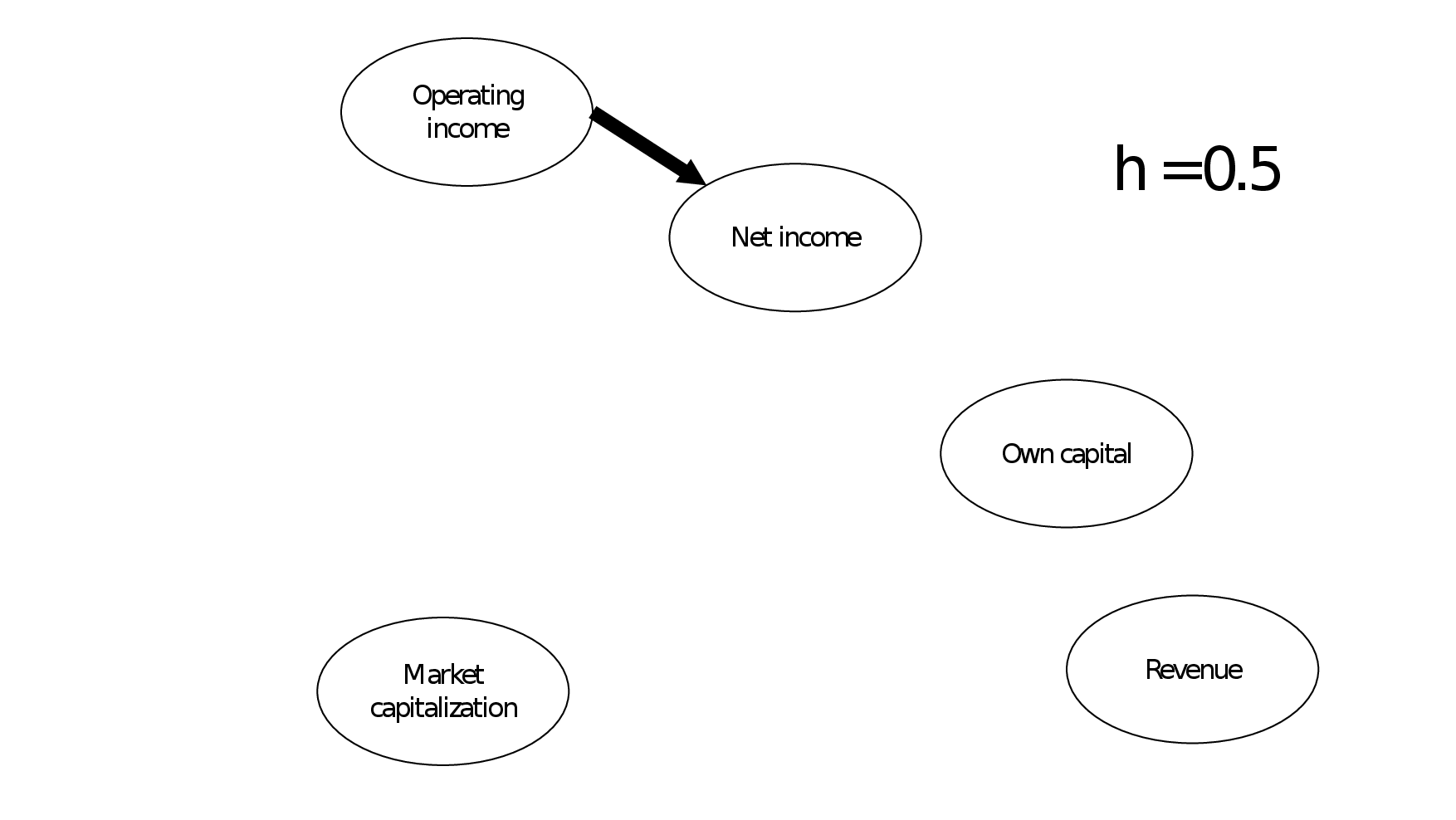}
}
\put(-170,75){(9)}
\end{center}
\end{minipage}\\

%\vspace{-4cm}
\begin{minipage}{0.5\hsize}
\begin{center}
\fbox{
\includegraphics[scale=0.2]{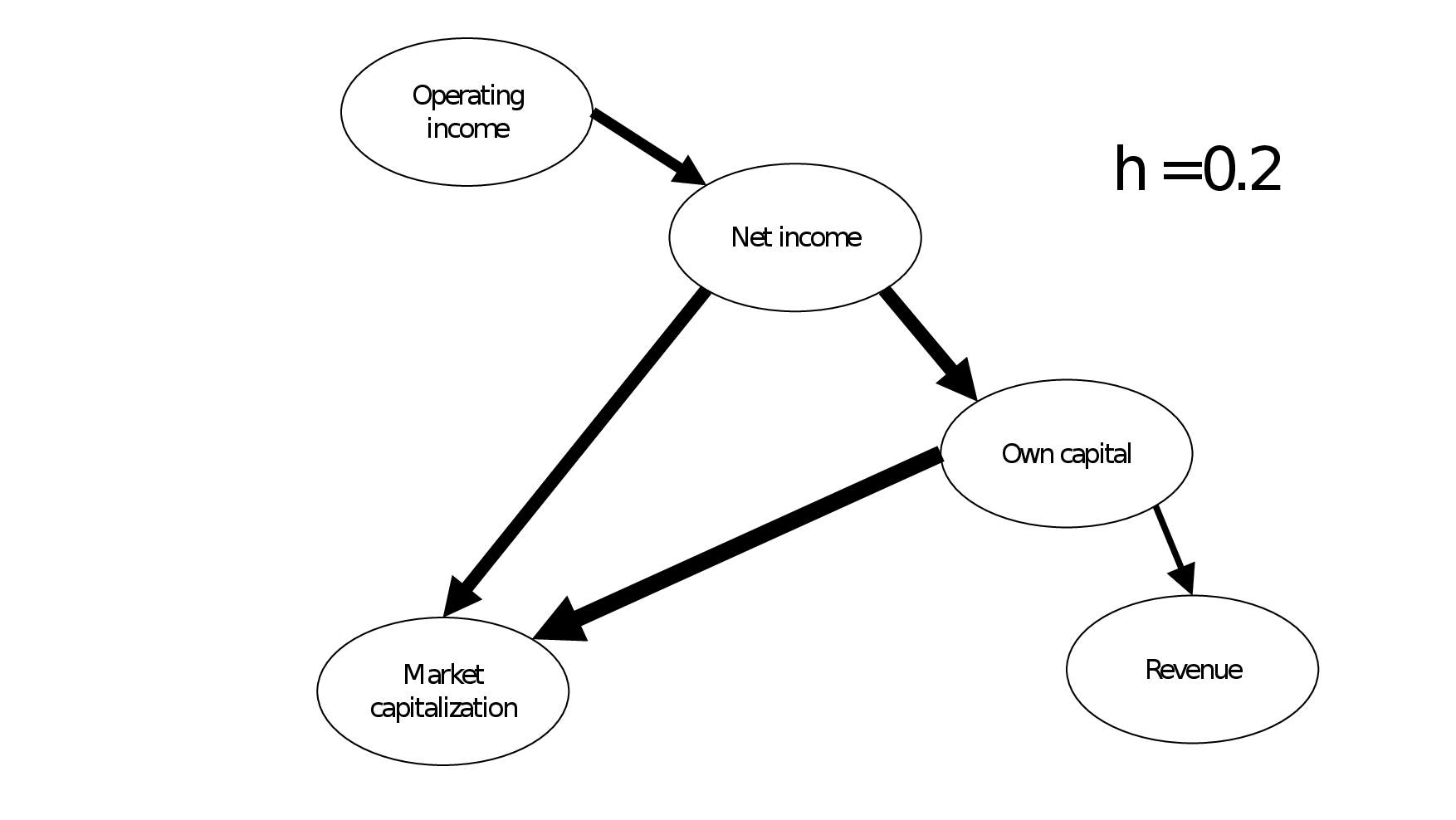}
}
\put(-170,75){(5)}
\end{center}
\end{minipage}
\begin{minipage}{0.5\hsize}
\begin{center}
\fbox{
\includegraphics[scale=0.2]{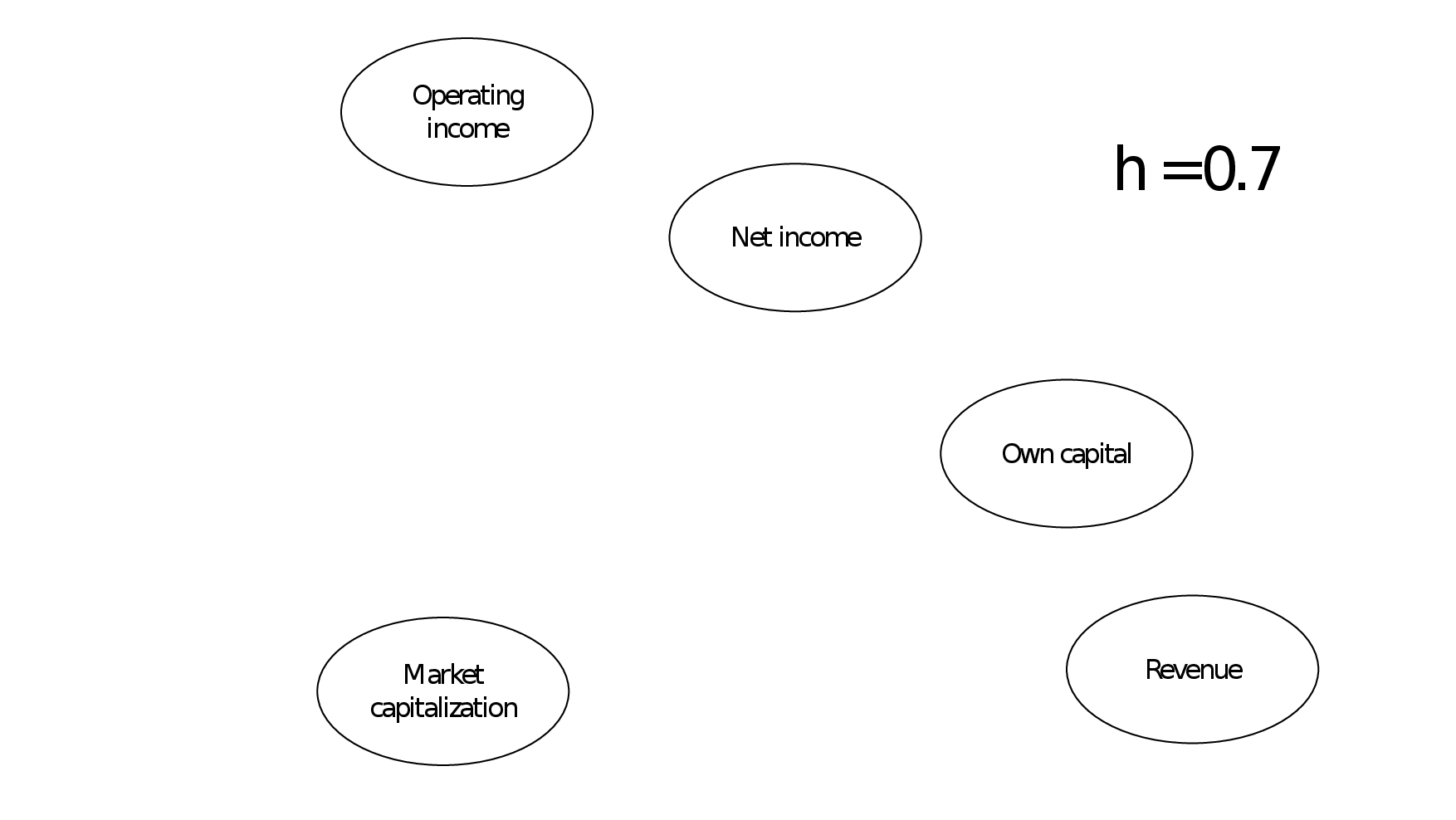}
}
\put(-170,75){(10)}
\end{center}
\end{minipage}

\end{tabular}

%\vspace{4cm}
\caption{Illustration of the changes in the directional network when the threshold $h$ is varied. The network of $h=0.2$ in (5), which takes the maximum number of links and thickness, is also displayed in Fig. \ref{fig:Directionality}.}
\label{fig:threshold and directional network}

\end{figure}

First, Fig. \ref{fig:Directionality} shows that operating income is a source node, while market capitalization and revenue are sinking nodes. In other words, operating income is the most influential accounting variable, while market capitalization and revenue are the most susceptible variables. Market capitalization is determined by the balance of supply and demand of market participants and the shareholders' strategy, which are mainly affected by the results of firms' operational management. Therefore, it is reasonable to assume that market capitalization is the most susceptible variable of the five variables.

Second, and more importantly, Fig. \ref{fig:Directionality} shows that net income and own capital influence market capitalization. This can be understood as follows. Market participants often evaluate the share price by the price--earnings ratio and price-to-book ratio computed from net income and own capital. Therefore, net income and own capital are thought to be the most influential items for market capitalization, as most investors often use these ratios as indicators. However, Fig. \ref{fig:Directionality} reveals that operating income has more influence on all accounting variables than net income and own capital, which are the only middle of influence flow in the directed network. This novel finding suggests that the usage of operating income rather than net income and own capital, which have been traditionally used to calculate the price--earnings ratio and the price-to-book ratio investment indicators, could lead to notably improved investment strategies.

Third, Fig. \ref{fig:Directionality} shows that operating income is the most influential item, affecting net income, which finally affects revenue. Revenue is the most susceptible item, which is against the calculation order of the statement of operation. Operatively, net income is calculated from operating income, which is calculated from revenue. Therefore, the order of influence in the context of income statement calculation is supposed to be in the order of revenue to operating income, and finally to net income. However, from our data-driven analysis, the influence order is different, which implies that operating income rather than revenue should be the focus of management strategy, which is our third main finding.

Since corporate finance and the structure of capital markets are common throughout the world, it is expected that we can observe a similar directionality network structure in other markets around the world. Indeed, it would be very interesting to extend this research methodology to the study of markets in other countries. In our view, the analysis could unveil not only the similarities but also the genuine differences of each market, providing a global understanding of the financial statement data and their interdependencies.

\section{Conclusion}

Previous studies of correlation coefficients in finance data have found out about the correlation between two variables but not their direction \cite{Preis,Plerou,Jiang,Podobnik,Podobnik2,Kumar}. However, we were able to capture the directionality between variables in financial statement data that cannot be captured by ordinary correlation coefficients.

In this study, the VC correlation approach unveiled the directionality between five major accounting variables, which is difficult to obtain using standard correlation methods. Our data-driven computations yielded new insights 
on major accounting variables, which can be translated into novel recommendations for
investment strategies. We summarize our findings as follows.

First, from the directionality network, we observed that operating income is the origin of influence on the other four accounting variables (net income, own capital, market capitalization, and revenue). Market capitalization and revenue are the most susceptible accounting variables. Second and more importantly, although market participants often focus on net income and own capitalization to evaluate the share price for investment strategy, operating income may be a better accounting variable on which to focus. Third, the influence order of revenue, operating income, and net income differs from the order of accounting calculation of income statements.

In summary, we believe that these results may lead to improved performance of financial management and application of optimal financial strategies for firms in future operations. Financial engineering-related areas, such as fraud detection and profitability analysis, could also benefit from our findings. In future work, we aim to expand the number of accounting variables to obtain a large-scale map of the directional interactions that governs global financial flows.

\section{Acknowledgments} 
T.O. was partially supported by a JSPS Grant-in-Aid for Scientific Research (Grant Number 15K01200).

\newpage

\begin{table}[h!]

\begin{center}

\begin{tabular}{|c|c|c|c|} % <-- Alignments: 1st column left, 2nd middle and 3rd right, with vertical lines in between

\hline

$[s,s^\prime]$ & $E_C$ & $\sigma_C$ & $N^\prime$ \\

\hline\hline

[net-income(i), market-cap(m)] & 0.308 & 0.219 & 1310  \\

\hline

[own-capital(o), market-cap(m)] & 0.337 & 0.232 & 1388 \\

\hline

[revenue(r), market-cap(m)] & 0.169 & 0.218 & 1184 
 \\

\hline

[operating-income(p), market-cap(m)] & 0.317& 0.206 & 1419  \\

\hline

[net-income(i), own-capital(o)] & 0.427& 0.216 & 1366 
 \\

\hline

[net-income(i), operating-income(p)] & 0.610& 0.268 & 1408  \\

\hline

[net-income(i), revenue(r)] & 0.273& 0.293 & 1149 
 \\

\hline

[operating-income(p), own-capital(o)] & 0.302& 0.221 & 1376 
 \\

\hline

[operating-income(p), revenue(r)] & 0.505& 0.278 & 1461 \\

\hline

[own-capital(o), revenue(r)] & 0.293& 0.257 & 1301  \\ \hline

\end{tabular}

\caption{Average of correlation coefficients  $E_C[s,s^\prime]$, standard deviation of correlation coefficients $\sigma_C{[s,s^\prime]}$, and number of firms $N^\prime$ for each pair of accounting variables $[s,s^\prime]$. }

\label{tab: correlation coefficients}

\end{center}

\end{table}

\begin{table}[h!]

\begin{center}

\begin{tabular}{|c|c|c|c|c|c|} % <-- Alignments: 1st column left, 2nd middle and 3rd right, with vertical lines in between

\hline

$[s,s^\prime]$ & $E_{\Delta F}$ & $\sigma_{\Delta F}$ & p-value & directionality \\

\hline\hline

[net-income(i), market-cap(m)] &  0.0110 & 0.0687 & $6.49\times10^{-9}$ & $\rightarrow$ \\

\hline

[own-capital(o), market-cap(m)] &  0.0129 & 0.0730 & $3.86\times10^{-11}$  &$\rightarrow$\\

\hline

[revenue(r), market-cap(m)] &  -0.000422 & 0.0502 & $ 0.772$ & \\

\hline

[operating-income(p), market-cap(m)] &  0.00299 & 0.0613 & $0.0657$& \\

\hline

[net-income(i), own-capital(o)] & 0.0139 & 0.0892 & $ 7.30\times10^{-9}$&$\rightarrow$ \\

\hline

[net-income(i), operating-income(p)] &  -0.00712 & 0.0529
 & $4.56\times10^{-7}$ &$\leftarrow$ \\

\hline

[net-income(i),revenue(r)] &  0.00441 & 0.0570 & $0.00858$ &\\ \hline

[operating-income(p), own-capital(o)] & -0.00217 &0.0664 & $0.225$ &\\

\hline

[operating-income(p), revenue(r)] & 0.00395 & 0.0499 & $0.00246$ &\\

\hline

[own-capital(o), revenue(r)] & 0.00734 & 0.0709
 & $0.000189$ &$\rightarrow$ \\ \hline

\end{tabular}

\caption{The average $E_{\Delta F}{[s,s^\prime]}$ and standard deviations $\sigma_{\Delta F}{[s,s^\prime]}$ of the difference in VC correlation, p-value, and directionality of each pair of accounting variables $[s,s^\prime]$ are shown. In the right-most column, $\rightarrow$ (resp. $\leftarrow$) indicates that the directionality is from $s$ to $s^\prime$ (resp. from $s^\prime$ to $s$). These directionalities are shown as an oriented network in Fig. \ref{fig:Directionality}.}

\label{tab: VC correlation}

\end{center}

\end{table}

\begin{table}[h!]
\begin{center}

\begin{tabular}{|c|c|c|c|c|c|} % <-- Alignments: 1st column left, 2nd middle and 3rd right, with vertical lines in between

\hline

$[s,s^\prime]$ & $E_{\Delta F}$ & $\sigma_{\Delta F}$ & p-value & directionality \\

\hline\hline

[net-income(i), market-cap(m)] (even) &  0.00964 & 0.0651 & $1.57\times10^{-4}$ & $\rightarrow$\\

\hline

[net-income(i), market-cap(m)] (odd) &  0.0123 & 0.0720 & $1.05\times10^{-5}$ & $\rightarrow$ \\

\hline

[own-capital(o), market-cap(m)] (even) &  0.0136 & 0.0760 & $2.68\times10^{-6}$ &$\rightarrow$\\

\hline

[own-capital(o), market-cap(m)] (odd) &  0.0123 & 0.0699 & $3.19\times10^{-6}$ &$\rightarrow$\\

\hline

[net-income(i), own-capital(o)] (even) &  0.0111 & 0.0866 & $9.11\times10^{-4}$&$\rightarrow$\\

\hline

[net-income(i), own-capital(o)] (odd) &  0.0167 & 0.0916 & $1.47\times10^{-6}$&$\rightarrow$ \\

\hline

[net-income(i), operating-income(p)] (even) &  -0.00794 & 0.0568 & $0.000222$&$\leftarrow$\\

\hline

[net-income(i), operating-income(p)] (odd) &  -0.00631 & 0.0488 & $0.000578$& ($\leftarrow$) \\

\hline

[own-capital(o), revenue(r)] (even) & 0.0102 & 0.0672 & $0.000125$ & $\rightarrow$\\ \hline

[own-capital(o), revenue(r)] (odd) & 0.00458 & 0.0741 & $0.109$ & ($\rightarrow$)\\ \hline

\end{tabular}

\caption{After dividing the data into two groups, the average $E_{\Delta F}{[s,s^\prime]}$ and standard deviations $\sigma_{\Delta F}{[s,s^\prime]}$ of the difference in VC correlation, p-value, and directionality of each pair of accounting variables $[s,s^\prime]$ are shown. In the right-most column, $\rightarrow$ (resp. $\leftarrow$) indicates that the directionality is from $s$ to $s^\prime$ (resp. from $s^\prime$ to $s$). The arrow enclosed in parentheses shows directionality with a statistically significant p-value threshold between $p_0$ and 0.2. All arrows are consistent with the original predictions in Table \ref{tab: VC correlation}.}

\label{tab: VC correlation validation splitting}

\end{center}

\end{table}

\begin{table}[h!]
\begin{center}
\begin{tabular}{|c|c|c|c|c|c|} % <-- Alignments: 1st column left, 2nd middle and 3rd right, with vertical lines in between

\hline

$[s,s^\prime]$ & $E_{\Delta F}$ & $\sigma_{\Delta F}$ & p-value & directionality \\

\hline\hline

[net-income(i), market-cap(m)] (r) &  $1.10\times10^{-3}$ & 0.0695 & 0.564 & \\

\hline

[own-capital(o), market-cap(m)] (r) &  $-5.31\times10^{-4}$ & 0.0741 & 0.789 &\\

\hline

[net-income(i), own-capital(o)] (r) &   $-1.42\times10^{-3}$ & 0.0903 & 0.560 & \\

\hline

[net-income(i), operating-income(p)] (r) &  $8.28\times10^{-4}$ & 0.0534 &0.560& \\

\hline

[own-capital(o), revenue(r)] (r) & $1.89\times10^{-3}$ & 0.0712 & $0.338$  & \\ \hline

\end{tabular}

\caption{After randomizing the data by exchanging the accounting variables of the pairs for each company, the average $E_{\Delta F}{[s,s^\prime]}$ and standard deviations $\sigma_{\Delta F}{[s,s^\prime]}$ of the difference in VC correlation, p-value, and directionality of each pair of accounting variables $[s,s^\prime]$ are shown. All p-values are greater than $p_0$, which implies that there is no statistical significance for directionality, as expected. In the right-most column, we show no directionality for all variable pairs.}

\label{tab: VC correlation validation randomized}

\end{center}

\end{table}
\newpage

\newpage

\newpage

\end{document}